\documentclass[iop,twocolappendix]{emulateapj}

\usepackage{color}
\usepackage{lineno}

\usepackage{graphicx}
\usepackage{float}
\usepackage{amsmath}
\usepackage{xcolor}

\usepackage[colorlinks=true,citecolor=blue,linkcolor=blue,urlcolor=blue]{hyperref}
\usepackage{enumitem}
\setlist{parsep=0pt,listparindent=\parindent}
    {\pagebreak[4]\global\pdfpageattr\expandafter{\the\pdfpageattr/Rotate 90}}%
    {\pagebreak[4]\global\pdfpageattr\expandafter{\the\pdfpageattr/Rotate 0}}%

\newcommand{\JHU}{Department of Physics and Astronomy, The Johns Hopkins University, Baltimore, MD 21218.}
\newcommand{\STScI}{Space Telescope Science Institute, Baltimore, MD 21218.}
\newcommand{\SAAO}{South African Astronomical Observatory, PO Box 9, 7935 Observatory, South Africa}
\newcommand{\UCT}{Department of Astronomy, University of Cape Town, 7701 Rondebosch, South Africa.}
\newcommand{\UCSC}{Department of Astronomy and Astrophysics, University of California, Santa Cruz, CA 92064, USA}
\newcommand{\AM}{George P. and Cynthia W. Mitchell Institute for Fundamental Physics \& Astronomy, Department of Physics \& Astronomy, Texas A\&M University, College Station, TX, USA}
\newcommand{\UCBerkeley}{Department of Astronomy, University of California, Berkeley, CA 94720-3411, USA}
\newcommand{\ESO}{European Southern Observatory, Karl-Schwarzschild-Str. 2, 85748 Garching b. M\"unchen, Germany}

\altaffiltext{1}{\JHU}
\altaffiltext{2}{\STScI}
\altaffiltext{3}{\UCBerkeley}
\altaffiltext{4}{\AM}
\altaffiltext{5}{\UCSC}
\altaffiltext{6}{\SAAO}
\altaffiltext{7}{\UCT}
\altaffiltext{8}{\ESO}

\newcommand{\intscatter}{0.14}
\newcommand{\ssample}{$3951$} 
\newcommand{\stetsoncut}{$L \geq 1.75$}

\newcommand{\baseline}{305 }
\newcommand{\apcor}{0.023}
\newcommand{\apcorerr}{+/- 0.01}

\newcommand{\goldsize}{139}
\newcommand{\silversize}{248}

\newcommand{\totalbronzesize}{438}
\newcommand{\totalsilversize}{296}
\newcommand{\totalgoldsize}{161}

\newcommand{\goldzpt}{23.24}
\newcommand{\silverzpt}{23.25}
\newcommand{\bronzezpt}{23.25}

\newcommand{\goldmod}{10.95}
\newcommand{\silvermod}{10.97}
\newcommand{\bronzemod}{10.97}

\newcommand{\goldrerr}{0.01}
\newcommand{\silverrerr}{0.01}
\newcommand{\bronzererr}{0.01}

\newcommand{\syserr}{0.06}
\newcommand{\silversyserr}{0.07}
\newcommand{\bronzesyserr}{0.08}

\newcommand{\colorcuterr}{0.024}
\newcommand{\jhcolorerr}{0.038}
\newcommand{\amperr}{0.042}
\newcommand{\scattererr}{0.021}

\newcommand{\goldabsmag}{-6.15 \pm 0.09}

\begin{document}

\title{A Near-Infrared Period-Luminosity Relation for Miras in NGC 4258, An Anchor for a New Distance Ladder}
\author{Caroline D. Huang \altaffilmark{1}, Adam G. Riess\altaffilmark{1,2}, Samantha L. Hoffmann \altaffilmark{2}, Christopher Klein\altaffilmark{3}, Joshua Bloom\altaffilmark{3}, Wenlong Yuan \altaffilmark{1,4}, Lucas M. Macri \altaffilmark{4}, David O. Jones\altaffilmark{1,5}, Patricia A. Whitelock \altaffilmark{6,7}, Stefano Casertano \altaffilmark{2}, Richard I. Anderson \altaffilmark{1,8}}

\begin{abstract}

We present year-long, near-infrared \emph{Hubble} Space Telescope WFC3 observations of Mira variables in the water megamaser host galaxy NGC 4258. Miras are AGB variables that can be divided into oxygen- (O-) and carbon- (C-) rich subclasses. Oxygen-rich Miras follow a tight (scatter $\sim 0.14$ mag) Period-Luminosity Relation (PLR) in the near-infrared and can be used to measure extragalactic distances. The water megamaser in NGC 4258 gives a geometric distance to the galaxy accurate to 2.6\% that can serve to calibrate the Mira PLR. We develop criteria for detecting and classifying O-rich Miras with optical and NIR data as well as NIR data alone. In total, we discover $\totalbronzesize$ Mira candidates that we classify with high confidence as O-rich. Our most stringent criteria produce a sample of $\goldsize $ Mira candidates that we use to measure a PLR. We use the OGLE-III sample of O-rich Miras in the LMC to obtain a relative distance modulus, $\mu_{4258} - \mu_{LMC} = \goldmod \pm \goldrerr$ (statistical) $\pm \syserr$ (systematic) mag which is statistically consistent with the relative distance determined using Cepheids. These results demonstrate the feasibility of discovering and characterizing Miras using the near-infrared with the \emph{Hubble} Space Telescope and the upcoming \emph{James Webb} Space Telescope and using them to measure extragalactic distances and determine the Hubble constant.  

\end{abstract}

\section{Introduction}
\label{sec:intro}

\begin{figure*}
\epsscale{1.2}
	\plotone{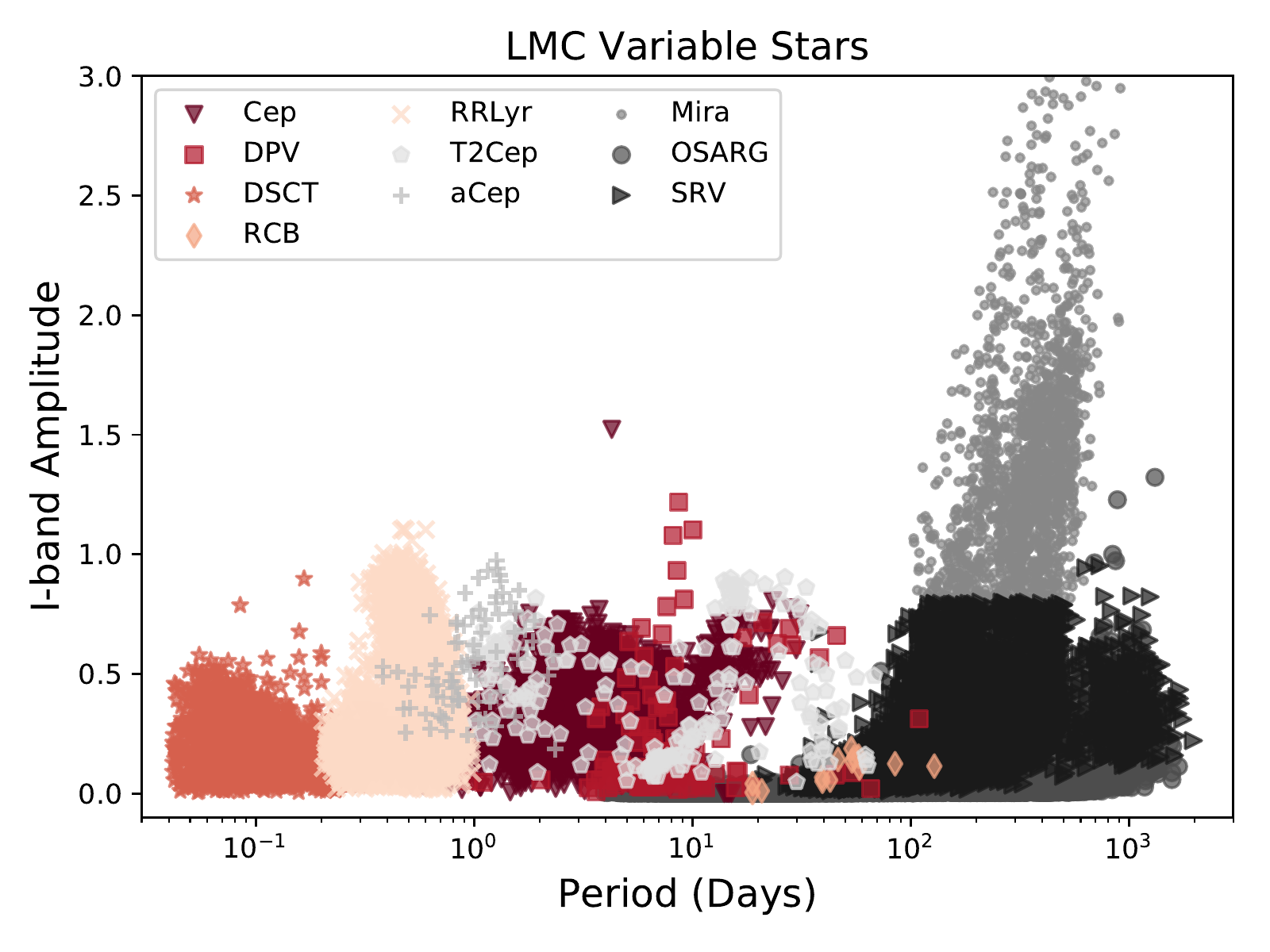}
	\caption{The distribution of I-band amplitude and period for LMC variable stars discovered by the Optical Gravitational Lensing Experiment (OGLE) III Survey \citep{Udalski08, Soszynski08a, Soszynski08b, Soszynski09a, Soszynski09b, Soszynski09c, Poleski10a, Poleski10b}. Abbreviations are as follows: Cep--Classical Cepheid; DPV--Double Period Variable; DSCT--$\delta$ Scuti variable; RCB--R Coronae Borealis variable; RRLyr--RR Lyrae; T2Cep--Type II Cepheid; 
	aCep--anomalous Cepheid; Mira--Mira; OSARG--OGLE Small Amplitude Red Giant; SRV--Semi-Regular Variable. 
	Miras and Semi-Regular Variables (SRVs) are separated on the basis of $I$ band amplitude of variation, but this distinction is somewhat arbitrary \citep{Soszynski05}. Miras, Cepheids, Type II Cepheids, RR Lyrae, and SRVs are radially-pulsating variables that follow Period-Luminosity Relations.}
	\label{fig:allvariablesiamp}

\end{figure*}

The value of the Hubble constant ($H_0$), the current expansion rate of the Universe, is a source of great interest in astrophysics. The improved precision in $H_0$ measurements \citep{Riess16} (hereafter, R16) has revealed a $3.4 \sigma$ discrepancy with the value inferred from observations of the cosmic microwave background (CMB) under the assumption of a $\Lambda$CDM cosmology \citep{Planck16XIII}. New parallax measurements of 7 long-period Cepheids in the Milky Way \citep{Riess18} combined with the R16 results increases the tension with Planck to 3.7 sigma. Although the local results have been confirmed \citep{Follin17, Dhawan17, Bonvin17} and the discrepancy is not dependent on any one datum \citep{Addison17}, the standard of proof is high for new physics and additional crosschecks are warranted. 

The most precise local measurement of $H_0$ relies on Cepheid variables as distance indicators R16 to calibrate the luminosity of type Ia supernovae (hereafter, SNe Ia) in hosts at nearby distances of 10-40 Mpc. Cepheids remain the best understood and most vetted primary distance indicator \citep{Freedman01, Bono10}. 

The next generation of space-based telescope, the \emph{James Webb} Space Telescope (\emph{JWST}), will not have the optical filters equivalent to those found in \emph{Hubble} Space Telescope (\emph{HST}), making it more difficult to search for Cepheids beyond $\sim 40$ Mpc to increase the sample of SNe Ia calibrators, which is essential to improve the precision of $H_0$. Cepheids are typically detected in the optical, where they have amplitudes $\sim 1$ mag. Their near-infrared (NIR) amplitude variations are much smaller ($\sim 0.3$ mag) and they lose their characteristic optical saw-toothed light curve shape which helps in their identification. This makes them more difficult to identify as variable stars at NIR wavelengths. As an alternative, \citet{Jang17} have used the Tip of the Red Giant Branch (TRGB) observed with \emph{HST} to check the Cepheid distances in nearby hosts, finding good agreement. However, because the TRGB is $\sim 2.5$ mag less luminous than Cepheids in the optical and $\sim 0.5$ mag less luminous in the NIR, this method will not be able to measure distances within the same volume as a Cepheid distance ladder.

\begin{figure*}
\epsscale{1.2}
	\plotone{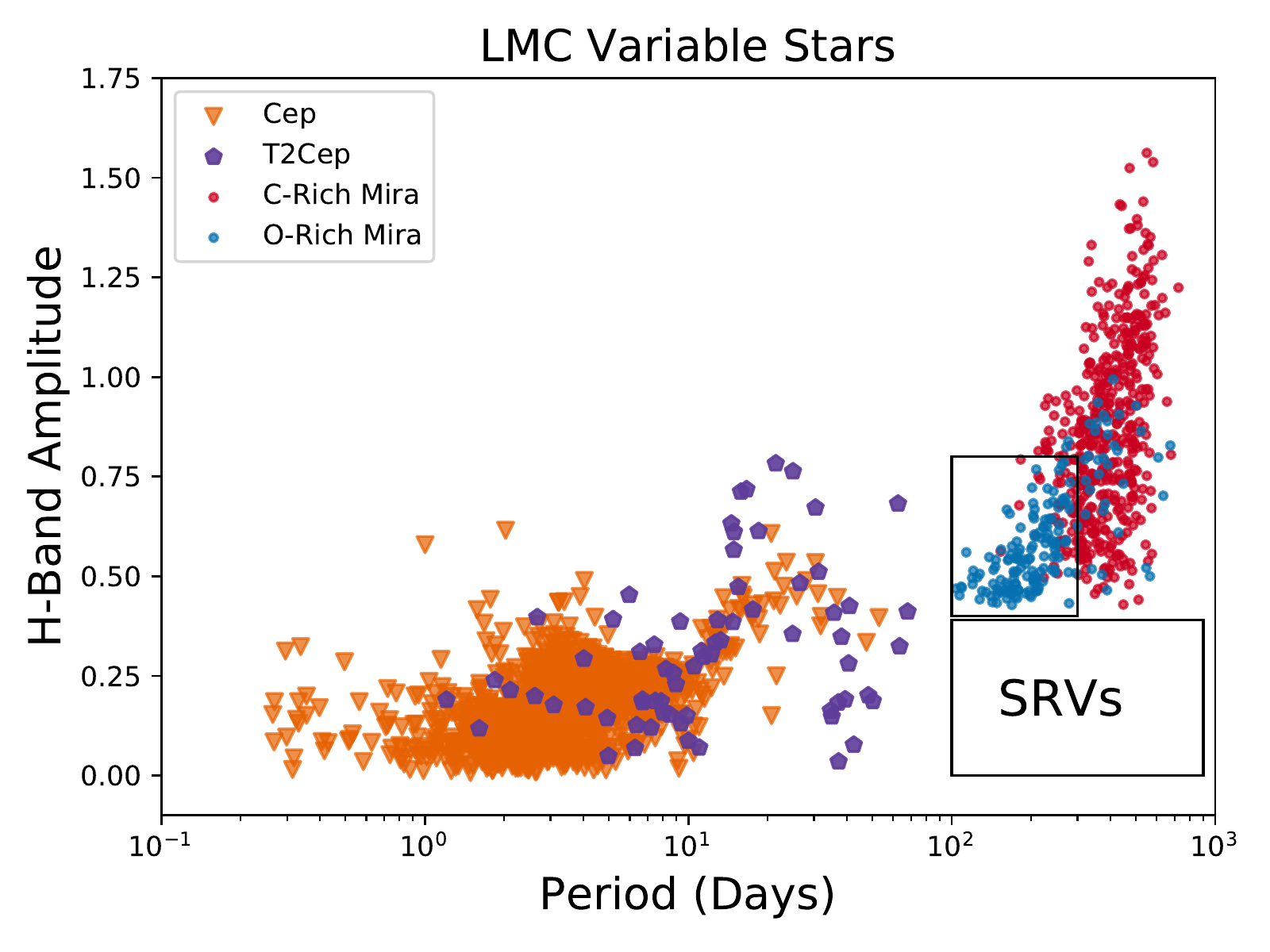}
	\caption{The distribution of $H$-band amplitude and period for LMC variable stars. Cepheid \citep{Macri15} and Type II Cepheid  \citep{Bhardwaj17} points use data from the LMC Near-Infrared Synoptic Survey. Abbreviations are same as for Figure \ref{fig:allvariablesiamp}. $H$-band amplitudes of Miras are from \citet{Yuan17b} and are estimated based on 3 NIR epochs. The boxes show the target selection of O-rich Miras and the location of SRVs. }
	\label{fig:allvariableshamp}
\end{figure*} 

A possible solution to this problem is using Mira variables to measure extragalactic distances. Miras are highly-evolved asymptotic giant branch (AGB) stars. They do not follow a tight PLR in the optical. However in the NIR, where they do follow a tight PLR, a 300-day Mira is roughly comparable in brightness to a 30-day Cepheid. They can provide an alternative distance indicator, allowing a simultaneous check of Cepheid distances and increasing the SNe Ia sample with \emph{JWST}. Miras are long-period ($P \gtrsim 100$ days), large amplitude ($\Delta V > 2.5$ mag, $\Delta I > 0.8$ mag) M or later spectral-type pulsating variable stars \citep{Kholopov85, Soszynski09b}. They are divided into oxygen- and carbon-rich spectral classes based on surface chemistry, with oxygen-rich (O-rich) Miras having a carbon-to-oxygen C/O ratio $< 1$ and a carbon-rich (C-rich) Miras having C/O $ > 1$. AGB stars with C/O $\sim 1$ are known as S stars. All stars are thought to enter the AGB phase as O-rich stars, but some evolve into C-rich stars due to dredge-up events \citep{Iben83}. The O-rich Miras have been shown to follow a tight ($\sigma \sim \intscatter$ mag) Period-Luminosity Relation (PLR) in the $K$-band \citep{Whitelock08, Yuan17a} that is comparable to the scatter in the Cepheid PLR in that bandpass \citep{Macri15}. The relation of Miras in the amplitude-period space relative to other classes of variables stars, including other distance indicators like Cepheids and RR Lyrae, is shown in Figure \ref{fig:allvariablesiamp}. As can be seen in the figure, they are distinguished from other variable stars by their large amplitudes and long, year-scale periods. 

Miras have a few advantages as distance indicators over Cepheids. Stars of a wide range of stellar masses go through the AGB phase, but Mira progenitors are typically of low-to-intermediate mass ($\sim 1 M_\odot$) while Cepheids have intermediate to high-mass progenitors ($ > 5 M_\odot$). Due to the bottom-heavy distribution of the stellar initial mass function (IMF), Mira progenitors are much more common than Cepheid progenitors. In addition, as seen in Figure \ref{fig:allvariableshamp}, their NIR amplitudes are about twice as large as NIR Cepheid amplitudes, making it much easier to discover these with infrared-only observatories like \emph{JWST}. Since they are older stars, Miras can also be found in galaxies without current star formation. This would allow the calibration of SNe Ia luminosities in early-type host galaxies with Miras. 

To test the efficacy of Miras as standard candles and demonstrate the feasibility of discovering and characterizing Miras with \emph{HST} and \emph{JWST}, we conducted a year-long, 12-epoch search for Mira variables in the megamaser-host galaxy NGC 4258 using WFC3 \emph{F160W} data. Our goals are to develop criteria for identifying and classifying Miras using primarily NIR photometry, to provide an initial test of the relative distances from Cepheids, and to obtain a calibrated Mira PLR relation relative to the Large Magellanic Cloud (LMC) by using the water megamaser distance to NGC 4258. 

Throughout the paper we refer to peak-to-trough variation in a Mira's magnitude over the course of one cycle as its `amplitude.' Statistical and systematic uncertainties are represented as $\sigma_r$ and $\sigma_{s}$ or with the subscripts $r$ and $s$ respectively. 

\begin{deluxetable*}{lcccccccc}
\tabletypesize{\scriptsize}
\tablecaption{{\it HST} Observations Used in this Work}
\tablewidth{0pt}
\tablehead{\colhead{Epoch} & \colhead{Proposal ID} & \colhead{Camera} & \colhead{UT Date} & \multicolumn{5}{c}{Exposure Time ($s$)} \\
\colhead{} & \colhead{} & \colhead{} & \colhead{} & \colhead{F435W} & \colhead{F555W} & \colhead{F814W} & \colhead{F125W} & \colhead{F160W}
}

\startdata

H-01 & 13445 & WFC3 & 2013-10-02 & \nodata & \nodata & \nodata & 2812 & 2812 \\
H-02 & 13445 & WFC3 & 2013-11-17 & \nodata & \nodata & \nodata & 353 & 2212 \\
H-03 & 13445 & WFC3 & 2013-12-03 & \nodata & \nodata & \nodata & 353 & 2212 \\
H-04 & 13445 & WFC3 & 2013-12-22 & \nodata & \nodata & \nodata & 353 & 2212 \\
H-05 & 13445 & WFC3 & 2014-01-26 & \nodata & \nodata & \nodata & 353 & 2212 \\
H-06 & 13445 & WFC3 & 2014-02-11 & \nodata & \nodata & \nodata & 353 & 2212 \\
H-07 & 13445 & WFC3 & 2014-03-14 & \nodata & \nodata & \nodata & 353 & 2212 \\
H-08 & 13445 & WFC3 & 2014-04-10 & \nodata & \nodata & \nodata & 353 & 2212 \\
H-09 & 13445 & WFC3 & 2014-05-11 & \nodata & \nodata & \nodata & 353 & 2212 \\
H-10 & 13445 & WFC3 & 2014-06-08 & \nodata & \nodata & \nodata & 353 & 2212 \\
H-11 & 13445 & WFC3 & 2014-07-05 & \nodata & \nodata & \nodata & 353 & 2212 \\
H-12 & 13445 & WFC3 & 2014-08-03 & \nodata & \nodata & \nodata & 353 & 2212 \\
O-01 & 9810 & ACS/WFC & 2003-12-06 & 1800 & 1600 & 800 & \nodata & \nodata \\
O-02 & 9810 & ACS/WFC & 2003-12-07 & 1800 & 1600 & 800 & \nodata & \nodata \\
O-03 & 9810 & ACS/WFC & 2003-12-08 & 1800 & 1600 & 800 & \nodata & \nodata\\
O-04 & 9810 & ACS/WFC & 2003-12-09 & 1800 & 1600 & 800 & \nodata & \nodata \\
O-05 & 9810 & ACS/WFC & 2003-12-11 & 1800 & 1600 & 800 & \nodata & \nodata \\
O-06 & 9810 & ACS/WFC & 2003-12-13 & 1800 & 1600 & 800 & \nodata & \nodata \\
O-07 & 9810 & ACS/WFC & 2003-12-16 & 1800 & 1600 & 800 & \nodata & \nodata \\
O-08 & 9810 & ACS/WFC & 2003-12-20 & 1800 & 1600 & 800 & \nodata & \nodata \\
O-09 & 9810 & ACS/WFC & 2003-12-24 & 1800 & 1600 & 800 & \nodata & \nodata \\
O-10 & 9810 & ACS/WFC & 2003-12-31 & 1800 & 1600 & 800 & \nodata & \nodata \\
O-11 & 9810 & ACS/WFC & 2004-01-08 & 1800 & 1600 & 800 & \nodata & \nodata \\
O-12 & 9810 & ACS/WFC & 2004-01-19 & 1800 & 1600 & 800 & \nodata & \nodata \\
M-13 & 11570 & WFC3 & 2009-12-17 & \nodata & \nodata & \nodata & \nodata & 2012 \\
M-14 & 11570 & WFC3 & 2009-12-17 & \nodata & \nodata & \nodata & \nodata & 2012 \\
M-13b & 11570 & WFC3 & 2010-05-29 & \nodata & \nodata & \nodata & \nodata & 2012 \\

\enddata
\tablecomments{Exposure times are rounded to the nearest second. The epochs labeled represent the results of three campaigns to observe NGC 4258. The H-xx epochs were aimed at discovering Miras. The O-xx epochs were observed previously to search for Cepheids. The M-xx epochs were observed in order to create a mosaic image of the whole galaxy. We have no epochs with simultaneous optical and infrared observations. } 
\label{tab:obs}
\end{deluxetable*}

\begin{figure*}[b]
\epsscale{1.2}
  \plotone{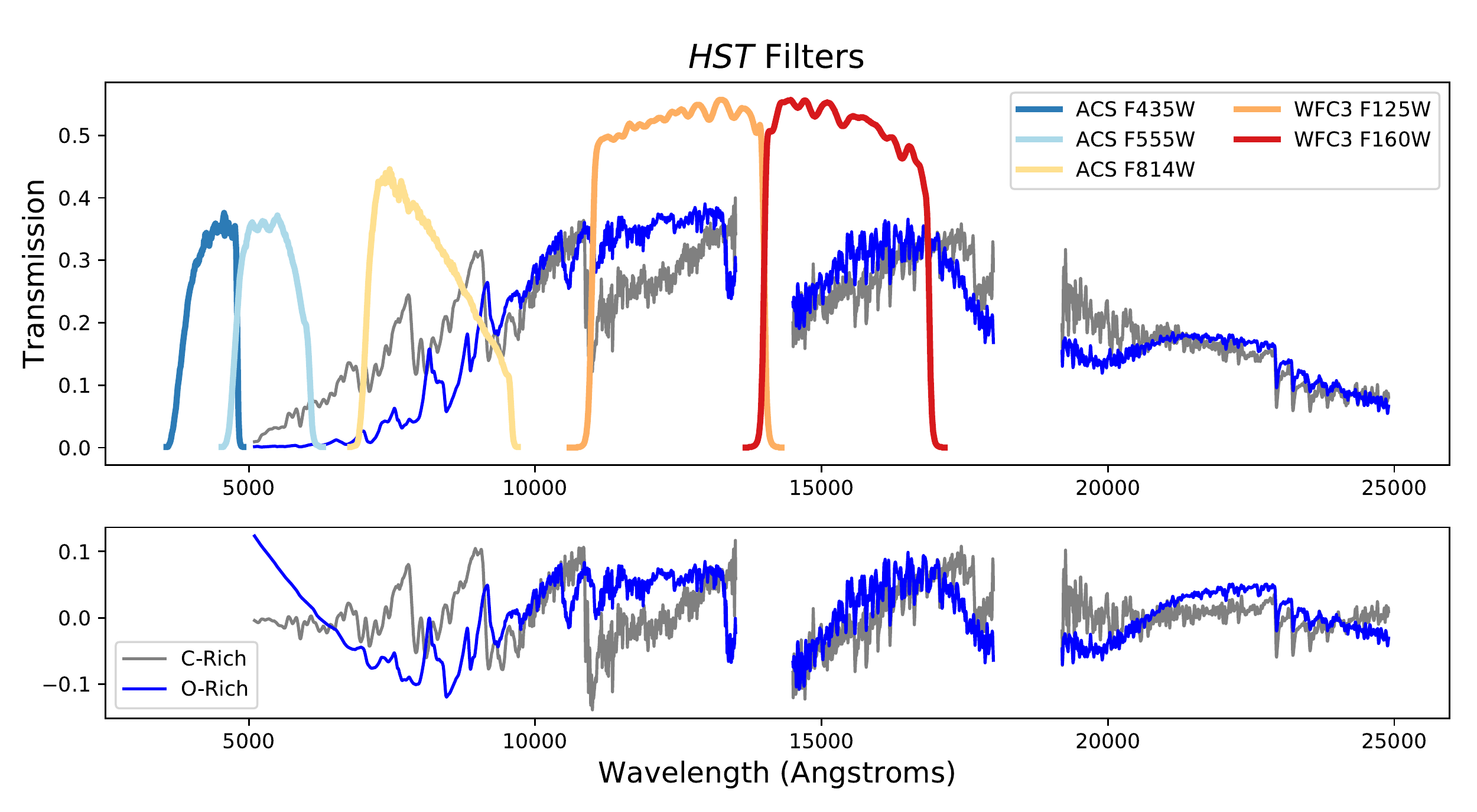}
  \caption{The transmission curves for every \emph{HST} filter in the images we analyzed shown with example O-rich and C-rich Mira spectra of the stars WW Sco and BH Cru respectively from \citet{Lancon00}. In the bottom image, the continuum emission has been subtracted out to better show the difference in spectral features. The \emph{F435W}, \emph{F555W}, \emph{F814W}, \emph{F125W}, \emph{F160W} filters roughly correspond to \emph{B, V, I, J, and H} in ground-based filter systems. 
  }
  \label{fig:filters}
\end{figure*}

\section{Observations, Data Reduction, and Photometry}
\label{sec:data}

\subsection{Observations}
\label{sec:obs}

With a roughly monthly observational cadence, we used twelve epochs of \emph{HST} WFC3 data (GO 13445; PI: Bloom) collected between October 2, 2013 and August 3, 2014 to search for Miras in NGC 4258. The field was chosen to overlap with the NGC 4258 ``inner" field from \citet{Macri06}, who discovered $\sim$ 50 Cepheids (see below for a description of these observations). The term ``inner" references the field's proximity to the nucleus of NGC 4258. The observations were centered at RA = $12^\text{h} 18^\text{m} 52^\text{s}.800$ and Dec = $+47^\circ 20' 19.70''$ (J2000.0). 

All twelve epochs have \emph{F160W} and \emph{F125W} data, but the first epoch is comprised of four $703s$ exposures in both the \emph{F160W} and \emph{F125W} (\emph{HST} $J$-band). The other epochs each contain four $553s$ \emph{F160W} exposures and one $353s$ \emph{F125W} exposure. 

This field was previously observed in \emph{F160W} (GO 11570; PI: Riess)
in December 2009 and May 2010. These earlier images were taken to create a mosaic of NGC 4258 in \emph{F160W} and to follow-up a number of long-period Cepheids discovered in the galaxy from the ground. This provided a thirteenth epoch of observation for most objects, while a few in the overlapping regions of the mosaic were observed twice and had fourteen epochs.

As already mentioned, this field was previously observed at optical wavelengths using \emph{HST} \citep[GO 9810; PI: Greenhill][]{Macri06}. These observations were taken with the Advanced Camera for Surveys (ACS) \citep{Ford03} in \emph{F435W}, \emph{F555W}, \emph{F814W}. The optical time-series consisted of twelve epochs between December 5, 2003 and January 19, 2004 with an observation spacing that followed a power law to allow for the detection of Cepheids at the largest possible range of periods. A summary of all of the observations used in the analysis is shown in Table \ref{tab:obs} and transmission curves for every filter are shown in Figure \ref{fig:filters}.

\subsection{Data Reduction}
\label{sec:reduc}

We use pipeline-processed images downloaded from the Canadian Astronomy Data Centre (CADC). For the \emph{F125W} and \emph{F160W} images taken between October 2013 and August 2014, we generated drizzled and stacked images for each epoch and filter using v.1.1.16 of Astrodrizzle \citep{Gonzaga12}. Each image contained four sub-pixel dither positions. The images were drizzled to a pixel scale of 0.08"/pix instead of the 0.13"/pix scale of WFC3 IR. As the baseline of observations of the field spanned almost a year, the roll angle of the camera changed by 282 degrees over the course of our observations. We choose the first epoch as the reference image and aligned all of the subsequent images onto it using DrizzlePac \citep{Gonzaga12}.

We used DAOMATCH and DAOMASTER, kindly provided by P. Stetson, to match sources in common between \citet{Macri06} and our \emph{F160W} master image. 

\subsection{Photometry and Calibration}
\label{sec:phot}

\setlength{\tabcolsep}{1em}
\begin{deluxetable*}{lccrrcc}
\tabletypesize{\scriptsize}
\tablecaption{Secondary Standards}
\tablewidth{0pt}
\tablehead{\colhead{ID} & \colhead{R.A.} & \colhead{Dec.} & \colhead{X} & \colhead{Y} & \colhead{F160W} & \colhead{Error} \\
\colhead{} & \colhead{(J2000.0)} & \colhead{(J2000.0)} & \colhead{(Pixels)} & \colhead{(Pixels)} & \colhead{(mag)} & \colhead{(mag)} 
}

\startdata

    49680 &   12 18 46.910   &     +47 19 58.43 &  987.352 & 1019.138 &   19.035 &    0.013 \\
    23564 &   12 18 48.964   &     +47 20 36.93 &  877.748 &  482.696 &   19.090 &    0.010 \\
     4663 &   12 18 50.894   &     +47 21 00.70 &  874.317 &   97.506 &   19.256 &    0.010 \\
    31187 &   12 18 51.819   &     +47 19 56.35 & 1481.836 &  637.801 &   19.298 &    0.011 \\
    35864 &   12 18 47.198   &     +47 20 25.75 &  795.737 &  734.066 &   19.309 &    0.012 \\
    51997 &   12 18 46.957   &     +47 19 53.17 & 1034.231 & 1065.619 &   19.389 &    0.010 \\
    44141 &   12 18 47.862   &     +47 20 02.36 & 1048.415 &  903.643 &   19.574 &    0.012 \\
     6657 &   12 18 50.314   &     +47 21 01.32 &  813.004 &  139.018 &   19.577 &    0.011 \\
    41446 &   12 18 47.622   &     +47 20 10.24 &  961.708 &  847.871 &   19.594 &    0.013 \\
     4234 &   12 18 51.396   &     +47 20 57.28 &  950.684 &   89.193 &   19.625 &    0.018 \\
     3655 &   12 18 51.193   &     +47 21 00.35 &  906.218 &   76.444 &   19.657 &    0.016 \\
     8400 &   12 18 50.221   &     +47 20 58.51 &  826.487 &  173.487 &   19.677 &    0.010 \\
    45991 &   12 18 46.740   &     +47 20 07.89 &  894.789 &  942.392 &   19.694 &    0.010 \\
     4662 &   12 18 50.548   &     +47 21 03.66 &  816.930 &   97.443 &   19.699 &    0.010 \\
    20550 &   12 18 48.903   &     +47 20 43.91 &  815.702 &  420.941 &   19.715 &    0.013 \\
    27518 &   12 18 48.969   &     +47 20 28.55 &  945.605 &  562.518 &   19.832 &    0.010 \\
    57524 &   12 18 48.801   &     +47 19 25.55 & 1435.620 & 1179.302 &   19.878 &    0.011 \\
    23751 &   12 18 48.816   &     +47 20 37.72 &  856.961 &  487.319 &   19.890 &    0.011 \\
    42544 &   12 18 48.425   &     +47 20 00.99 & 1114.240 &  870.805 &   19.913 &    0.013 \\
     7460 &   12 18 49.568   &     +47 21 06.00 &  702.760 &  155.089 &   19.918 &    0.011 \\
     3896 &   12 18 50.963   &     +47 21 01.74 &  872.693 &   81.896 &   19.925 &    0.016 \\
    10561 &   12 18 49.443   &     +47 21 00.47 &  735.012 &  218.259 &   19.937 &    0.012 \\
     4566 &   12 18 50.618   &     +47 21 03.22 &  827.237 &   95.984 &   19.991 &    0.009 \\
    22541 &   12 18 49.572   &     +47 20 33.95 &  960.815 &  461.519 &   20.052 &    0.013 \\
    14188 &   12 18 50.626   &     +47 20 42.79 &  992.267 &  290.814 &   20.057 &    0.011 \\
      695 &   12 18 52.032   &     +47 20 59.76 &  992.657 &   13.522 &   20.060 &    0.011 \\
    45068 &   12 18 48.285   &     +47 19 56.76 & 1134.596 &  922.657 &   20.150 &    0.012 \\
     8203 &   12 18 49.034   &     +47 21 09.03 &  626.493 &  169.699 &   20.212 &    0.009 \\
    42205 &   12 18 48.214   &     +47 20 03.52 & 1073.341 &  863.826 &   20.237 &    0.016 \\
    65132 &   12 18 45.898   &     +47 19 33.70 & 1087.567 & 1338.530 &   20.275 &    0.010 \\
    12889 &   12 18 50.029   &     +47 20 50.58 &  871.554 &  265.064 &   20.293 &    0.010 \\
    36017 &   12 18 48.265   &     +47 20 16.35 &  975.233 &  736.852 &   20.295 &    0.010 \\
    15965 &   12 18 48.742   &     +47 20 55.00 &  710.891 &  327.951 &   20.448 &    0.012 \\
     2890 &   12 18 51.222   &     +47 21 01.68 &  898.406 &   61.263 &   20.467 &    0.010 \\
    57706 &   12 18 49.091   &     +47 19 22.73 & 1486.569 & 1182.616 &   20.514 &    0.013 \\
    40960 &   12 18 44.791   &     +47 20 35.40 &  483.976 &  838.315 &   20.575 &    0.009 \\
    59832 &   12 18 44.802   &     +47 19 54.64 &  812.577 & 1227.611 &   20.609 &    0.009 \\
    59266 &   12 18 48.736   &     +47 19 22.40 & 1454.737 & 1214.741 &   20.642 &    0.010 \\
     4044 &   12 18 50.747   &     +47 21 03.30 &  839.175 &   84.681 &   20.673 &    0.010 \\
    48800 &   12 18 46.408   &     +47 20 04.56 &  889.228 & 1001.389 &   20.750 &    0.011 \\
    36245 &   12 18 45.236   &     +47 20 41.70 &  476.671 &  741.707 &   20.830 &    0.015 \\
    64758 &   12 18 46.901   &     +47 19 25.91 & 1247.840 & 1331.200 &   20.886 &    0.010 \\
    43584 &   12 18 43.617   &     +47 20 39.71 &  335.074 &  892.949 &   21.029 &    0.010 \\
    52996 &   12 18 45.086   &     +47 20 06.86 &  742.139 & 1087.443 &   21.039 &    0.011 \\

\enddata
\tablecomments{A list of all the secondary sources used calibrate the \emph{F160W} light curves. ID numbers are photometry IDs, X and Y positions are relative to the first epoch of the \emph{F160W} image, and the errors are photometric errors as estimated by DAOPHOT.} 
\label{tab:secondarystandards}
\end{deluxetable*}

Given that our fields are quite crowded, we used tools specifically designed for crowded fields: DAOPHOT/ALLSTAR \citep{Stetson87} and ALLFRAME \citep{Stetson94}. Our DAOPHOT procedure is different from the R16 NIR forced photometry because we conduct a search for Miras in WFC3 \emph{F160W} and do not know their positions \emph{a priori}. We created point-spread-functions (PSFs) in DAOPHOT with \emph{F160W} and \emph{F125W} exposures of the standard star P330E. Due to the crowded nature of our fields, we were unable to find enough isolated stars to make PSFs using sources in our field. Aperture corrections, discussed in \S \ref{sec:apcor}, were used to account for imperfections in the PSF model. 

We stacked all of the \emph{F160W} observations to make a deeper ``master" image. Then we used the DAOPHOT routine FIND to detect sources with a greater than $3 \sigma$ significance level in standard deviations from the sky background noise. Then the DAOPHOT routine PHOT was used to perform aperture photometry. We input the star list produced from the aperture photometry into ALLSTAR for PSF photometry, optimized for a 2.5 pixel full-width at half-maximum (FWHM). We then repeated these steps on the star-subtracted image generated by ALLSTAR (with all of the previously-discovered sources removed) to produce a second source list. The two source lists were then concatenated to create a master source list of $\sim 1.3 \times 10^5$ entries. 

We input this master source list into ALLFRAME. ALLFRAME is similar to ALLSTAR, except that it is capable of performing simultaneous fits to the profiles of all of the stars contained in all of the images of the same field. It then produces time-series PSF photometry as output. We used the master source list as the input for fitting all of the \emph{F160W} epochs at the same time.

We then searched for secondary standards in the star lists by choosing bright objects that had been observed in all twelve of the last \emph{F160W} epochs. We visually inspected the stellar profiles and their surroundings to choose secondary standards that were relatively isolated compared to other bright stars, removing any that showed variability or had large photometric errors. This left us with a total of 44 sources, summarized in Table \ref{tab:secondarystandards}. We calculated the celestial coordinates for all of these sources using the astrometric solutions in the FITS headers and PyAstronomy program pyasl. The mean residuals for these stars across all epochs of \emph{F160W} imaging exhibited a dispersion of 0.01 mag which is corrected for during the ALLFRAME photometry. 

\subsection{Aperture Corrections}
\label{sec:apcor}

We use aperture corrections to account for missing flux from imperfections in our PSF model. Using the standard stars chosen for the variability search, we subtract everything but these sources from the master image using the master source list with DAOPHOT's SUB task. The standard stars are already relatively bright and isolated for our field, but this subtraction helps to ensure that we remove any additional flux from the wings of the standard stars. We then perform aperture photometry on the sources using increasing aperture radii (up to $0.4$'') and check that the growth curves look well behaved over a range of apertures.

We calculated the difference between PSF and $0.4$'' aperture magnitudes, to which we added the flux beyond this limit previously calculated by STScI. The WFC3 \emph{F160W} Vegamag zeropoint was 24.5037. The overall correction from PSF to ``infinite aperture" magnitudes was \apcor \ \apcorerr \ mag. 

\section{Mira Selection Criteria}
\label{sec:selection}

\begin{figure*}
\epsscale{1.2}
  \plotone{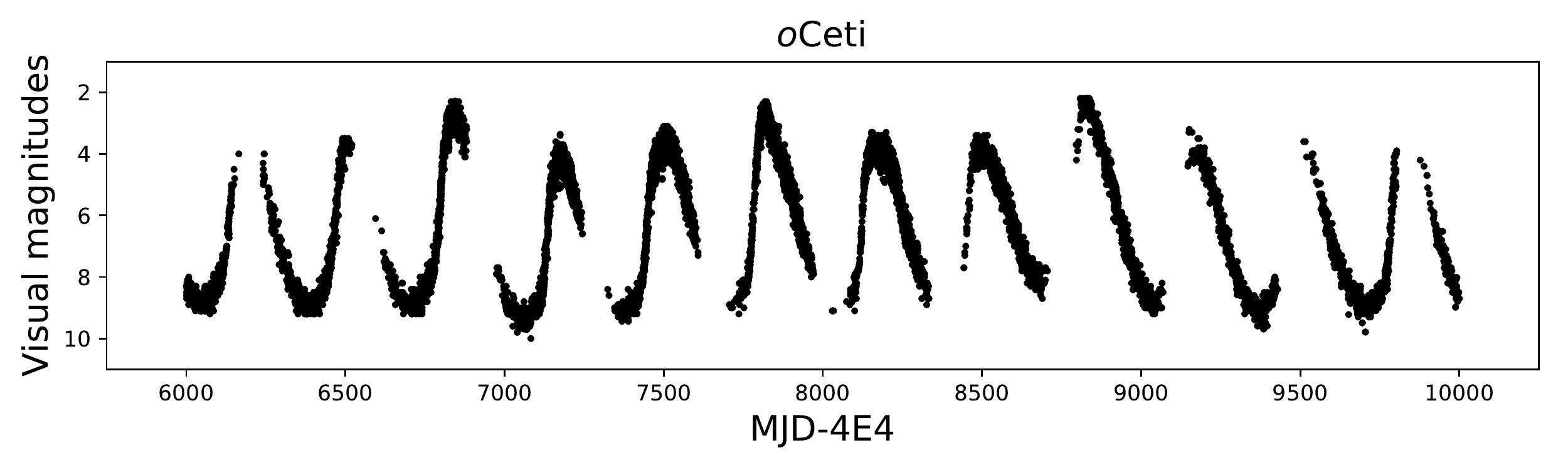}
  \caption{Visual light curve of $o$Ceti (Mira) using data from the American Association of Variable Star Observers (AAVSO). The light curve shows cycle-to-cycle variations in both amplitude and mean magnitude.} 
  \label{fig:MiraLC}
\end{figure*}

RR Lyrae, Classical Cepheids, and Type II Cepheids, among other variable stars, can be identified with the use of light curve templates, which exist for both the optical and near-infrared \citep{Jones96, Yoachim09, Sesar10, Inno15, Bhardwaj17} since the shapes of their light curves vary as function of period in a predictable way. The templates allow them to be identified as a particular class of variable star by light curve morphology after they have been initially identified as variable. However, Miras have irregular light curve shapes that are not strictly a function of period. They may also also exhibit variations in light curve shape or amplitude between different cycles. These cycle-to-cycle variations can be seen in the visual light curve of the prototype Mira, $o$Ceti, shown in Figure \ref{fig:MiraLC}. Figures 7 and 9 of \citet{Whitelock94} and Figure 1 of \citet{Olivier01} show the smaller cycle-to-cycle variations present in $K$-band Mira light curves. 

Miras and other long-period variables (LPVs) can also have a longer, secondary pulsational period in addition to their primary pulsational periods. These long secondary periods \citep[LSPs;][]{PayneGaposchkin54, Houk63, Nicholls09} are typically about an order of magnitude longer than the primary periods \citep{Wood99} and are found in about 25-50 percent of all LPVs \citep{Wood99, Percy04, Soszynski07b, Fraser08}. Several theories have been proposed to explain LSPs, but there is no general consensus \citep{Soszynski07b, Saio15, Percy16}. Both LSPs and cycle-to-cycle variations will result in the same Mira having different magnitudes at the same phase in different cycles. 
 
Instead of using template-fitting, Mira variables are typically identified only by their large $V$ and $I$ amplitudes and long periods. The amplitude criterion is used to separate Miras from the more numerous, lower-amplitude, and sometimes more inconsistent semi-regular variables (SRVs), which are another type of LPV (see Figures \ref{fig:allvariablesiamp} and \ref{fig:allvariableshamp}). The amplitude cutoffs have traditionally been defined in the optical as $\Delta V > 2.5$ mag or $\Delta I > 0.8$ mag \citep{Kholopov85, Soszynski09b} and the periods range from 80-1000 days, though there are some Miras with periods that can be significantly longer. A few previous studies also identified Miras in the infrared by using $J$, $H$, and $K$ time-series data \citep{Whitelock06, Whitelock09, Matsunaga09} but large amplitudes in the NIR do not always correspond to large amplitudes in the visible bands. A sample of Miras selected based on large NIR amplitudes could contain objects that would not be selected based on visual criteria and vice versa. 

\citet{Yuan17b} sought to differentiate between these classes of variable stars by using a Random-Forest classifier which incorporated period, light curve shape (O-rich Miras have more symmetric light curves), and other properties into the classification. However, our small number of epochs limits us to using simple cuts. Since we have only limited optical data (at most spanning $\sim$ 40\% of a Mira cycle due to the short baseline optimized to detect Cepheids) and only one band in the NIR with time-series photometry, we need selection criteria which rely more heavily on NIR measurements. 

We also note that not all Miras are good distance indicators. The two C-rich and O-rich subgroups follow different PLRs in the $H$-band \citep{Ita04, Ita11}. C-rich stars can also develop optically thick circumstellar dust shells that result in them appearing fainter even in the NIR \citep{Yuan17b}. Therefore, we use only O-rich Miras as distance indicators and must separate them from C-rich Miras. 

The photospheric differences between C- and O-rich Miras are thought to be due to the differing ratios of carbon and oxygen in their atmospheres. The carbon and oxygen in a star's atmosphere will combine to make the stable CO molecule until there is no more of the less abundant element. The excess carbon or oxygen will then be left over for dust formation and will combine to create molecules such as TiO and VO in O-rich stars or CN and C$_2$ in C-rich stars \citep{Cioni01}. These molecules define each spectral type and can also change a Mira's color, which has potential consequences that are discussed in greater detail in \S \ref{sec:jhcol}. While there are generally $J-K$ distinctions in color between C- and O-rich stars, the cutoff in color varies in different galaxies \citep{Cioni03a}. In addition, some O-rich stars may be very red, as in the case of OH/IR stars. 

We expect to encounter a smaller ratio of C-rich Miras to O-rich Miras in NGC 4258 than in the LMC. Higher ratios of O-rich to C-rich stars are observed in galaxies with higher metallicity \citep{Blanco83, Mouhcine03, Hamren15}. The inner field of NGC 4258 is expected to be $\sim 0.1$ dex more metal-rich than the LMC, though still $\sim -0.2$ dex relative to solar \citep{Bresolin11}. 

\subsection{Detection of Variability}
\label{sec:variability}

\begin{figure}
\epsscale{1.2}
  \plotone{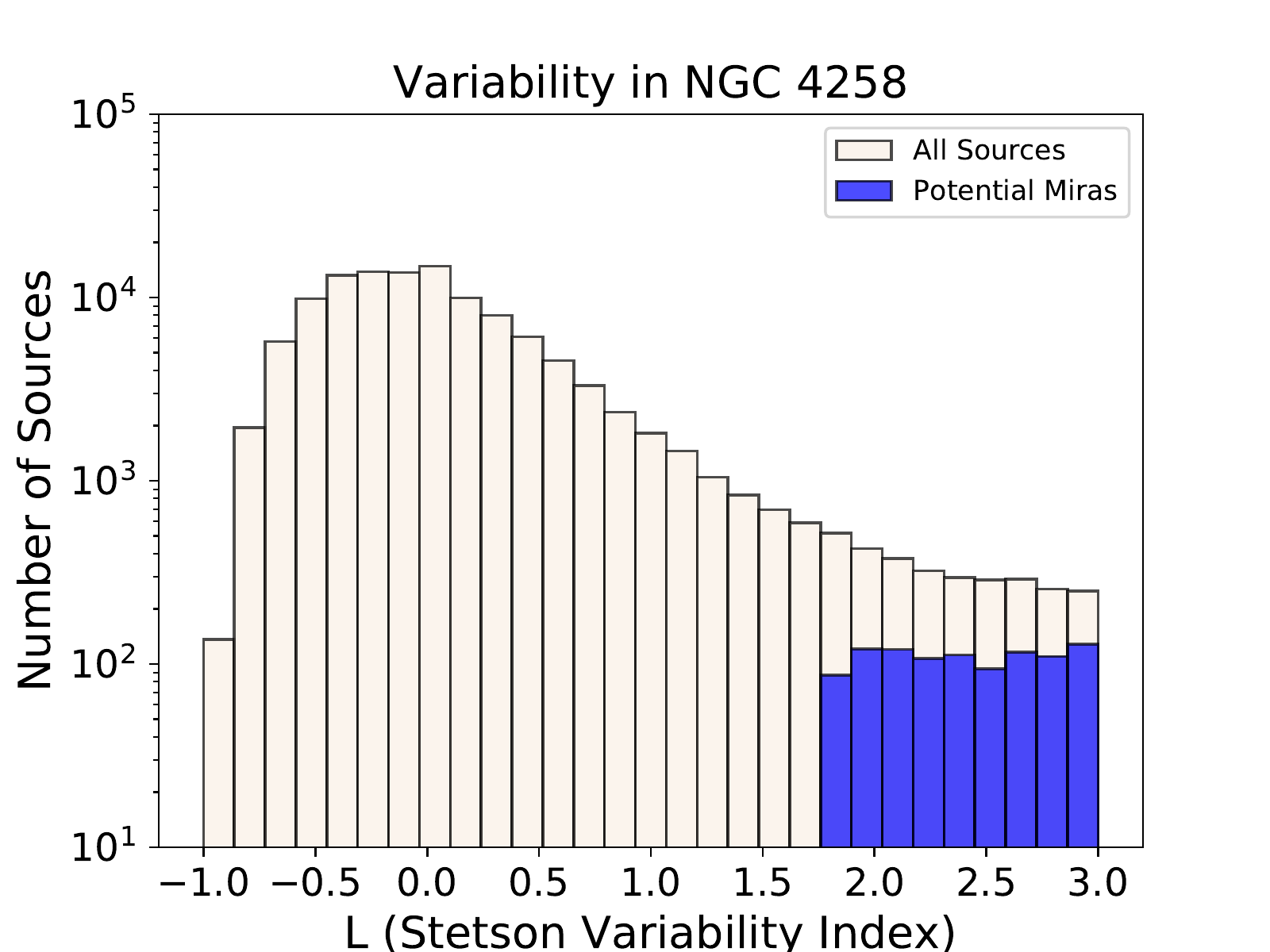}
  \caption{The distribution of the Welch-Stetson variability index $L$ for all of the sources in the field (shown in white) and for the subset of objects that made it past the initial visual inspection (in blue). All objects with \stetsoncut \, were visually examined and anything that did not pass visual inspection or had $\Delta \emph{F160W} < 0.4$ was automatically discarded.}
  \label{fig:variability}
\end{figure}

\begin{figure}
\epsscale{1.2}
  \plotone{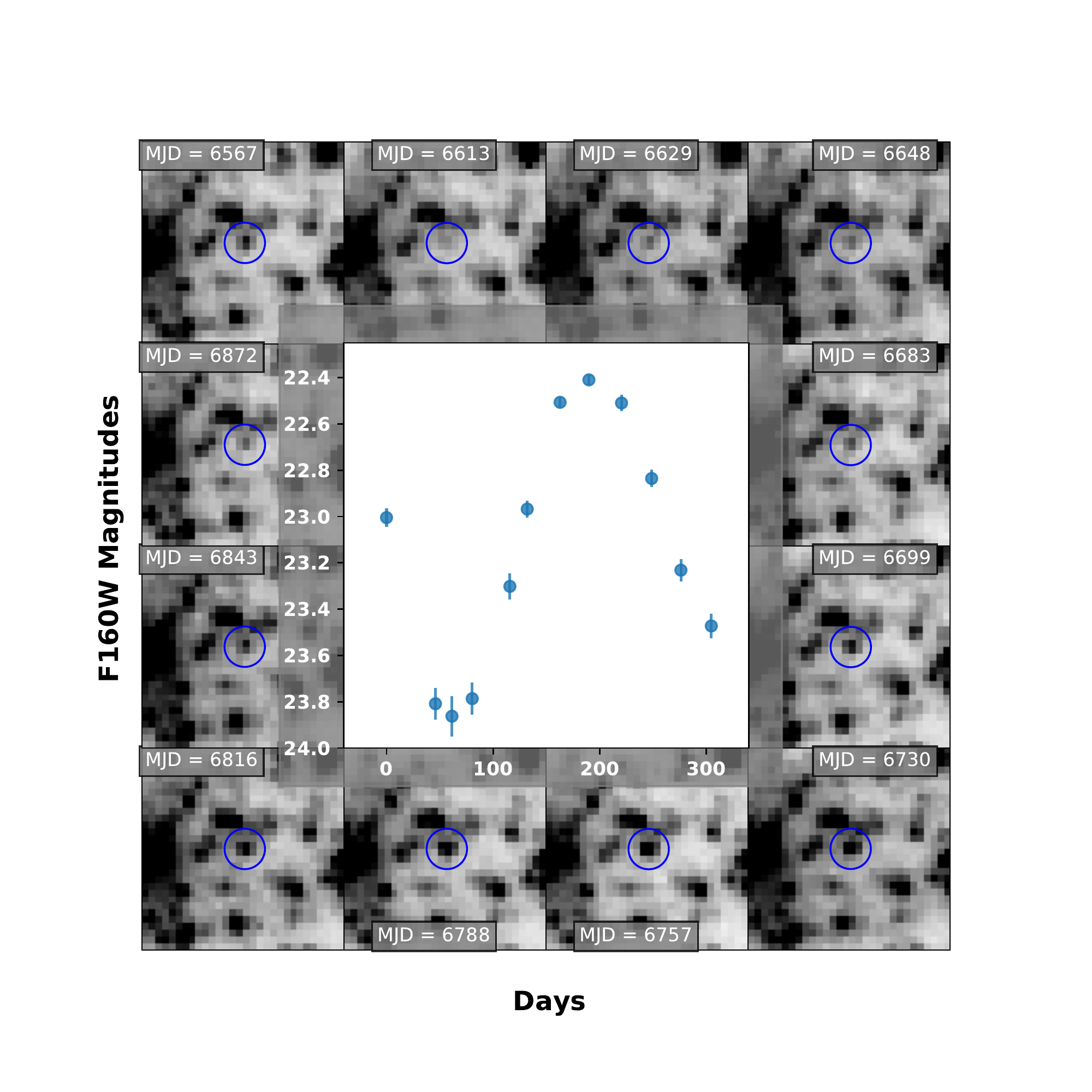}
  \caption{Light curve and finding charts for a 266 day Mira from NGC 4258. The Mira is the source in the blue ring in each of the finding charts. Numbers on each stamp represent the date of the observation, corresponding to MJD - 50000. The vertical axis of the inset plot gives the \emph{F160W} magnitude and the horizontal axis marks the time since the first epoch of observations in days. Each stamp corresponds to one point on the light curve, starting in the upper left corner and going clockwise around the image. As the Mira progresses through its light cycle, the brightness of the sources in the postage stamps noticeably changes. 
  }
  \label{fig:miralc}
\end{figure}

We first used the Welch-Stetson variability index \emph{L} \citep{Stetson96} to identify a sample of variable objects detected in all of the \emph{F160W} epochs. This is a combination of two other measures of variability (all three are defined in Equations 1, 2, and 3 of \citet{Stetson96}). Objects that have larger variances from their mean magnitudes, exhibit similar variability in multiple images taken at around the same time, and have non-Gaussian magnitude distribution will have a larger \emph{L} value. We calculated $L$ for every object identified in the master photometry list and then further inspected a number of objects that met our threshold of $L > 1.75$, about $3\sigma$ above the mean $L$ for all objects.

Figure \ref{fig:variability} shows the distribution of \emph{L} values for candidate  Miras and all sources in NGC 4258. We kept only sources with \stetsoncut \,that were detected in all twelve epochs of the most recent \emph{F160W} observations. Sources that did not show periodicity (were only continuously rising or decreasing light curves), had $\Delta F160W < 0.4$ or periods of less than 100 days were then removed from the list of possible Miras. The $ \Delta F160W > 0.4$ cut roughly corresponds to the $\Delta I > 0.8$ used by \citet{Soszynski09b} to distinguish between Miras and SRVs. Additional discussion of the appropriate minimum amplitude cut can be found in \S \ref{sec:hamp}. These requirements resulted in \ssample \, Mira candidates remaining in our sample. An example of an \emph{F160W} Mira light curve and its finding charts is shown in Figure \ref{fig:miralc}.

\subsection{Estimating Cycle-to-Cycle Variation}
\label{sec:cyclecyclevar}

\begin{figure}
\epsscale{1.2}
  \plotone{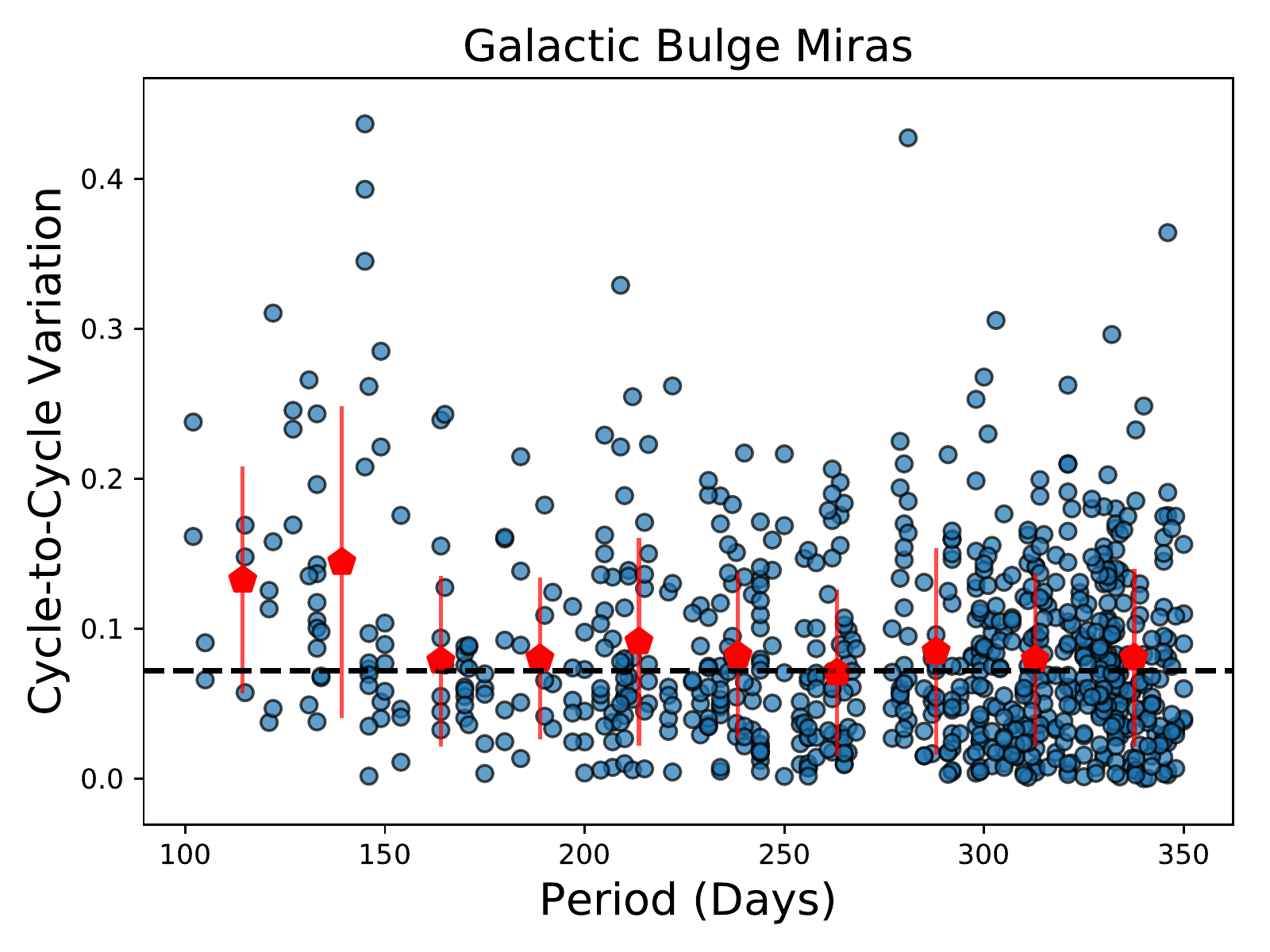}
  \caption{Distribution of cycle-to-cycle variations in Galactic Bulge Miras in the H-band using data from \citet{Matsunaga09}. Blue points show the cycle-to-cycle variation for each Mira studied, as a function of the Mira's period, red points are the means of each period bin along with the standard deviations of each bin. The bins illustrate the difference in the average cycle-to-cycle variation for Miras of different periods, but only the median of the cycle-to-cycle variations for all Miras was used as an additional error when fitting periods.} 
  \label{fig:cyclevar}
\end{figure}

We estimated the level of cycle-to-cycle variations in the $H$-band mean magnitude we might measure using data from \citet{Matsunaga09}, which looked at Miras in the Galactic Bulge. These Miras were observed in twelve fields in the NIR between 2001 and 2008, with observations of the same phases during various cycles each Mira. While most of our observations took place over a span of only $\sim 300$ days, we still need to account for possible variations in magnitude at a given phase between the main \emph{F160W}/$H$ band campaign in 2013 and 2014 and the observations taken in 2009, many oscillation cycles earlier. 

The \citet{Matsunaga09} dataset covers different objects than the OGLE-III dataset and is more sparsely sampled, but it is one of the largest sets of time-series observations of Miras in the NIR. Some of the OGLE-III Miras have been observed in the NIR as well as the optical bands used by OGLE, but the vast majority of the OGLE-III Miras have only a few or single epochs in the NIR. Because of this, we did not use the OGLE-III dataset to estimate NIR cycle-to-cycle variations.

We binned the Galactic Bulge data of each Mira candidate observed by \citet{Matsunaga09} into 30-day bins (to simulate the frequency of our observations) and calculated the $H$-band mean magnitude in each bin. Next we folded the binned magnitudes by the periods measured in \citet{Matsunaga09} to obtain their phases. We then binned the resulting points by phase to see how bright each Mira was at that particular phase over different cycles. Finally, we calculated the variance of the points in each phase bin to estimate the cycle-to-cycle variations in magnitude at similar phase for the observations. Figure \ref{fig:cyclevar} shows the cycle-to-cycle variation of mean magnitudes as a function of Mira period. As expected, the shorter-period Miras have larger `cycle-to-cycle' variations because each 30-day bin averages over a larger range of phases. The median cycle-to-cycle variation overall was 0.072 magnitudes, which we incorporated as an additional error added in quadrature when fitting periods using data from different cycles. 

\subsection{Determination of Periods}
\label{sec:periods}

\begin{figure}
\epsscale{1.2}
  \plotone{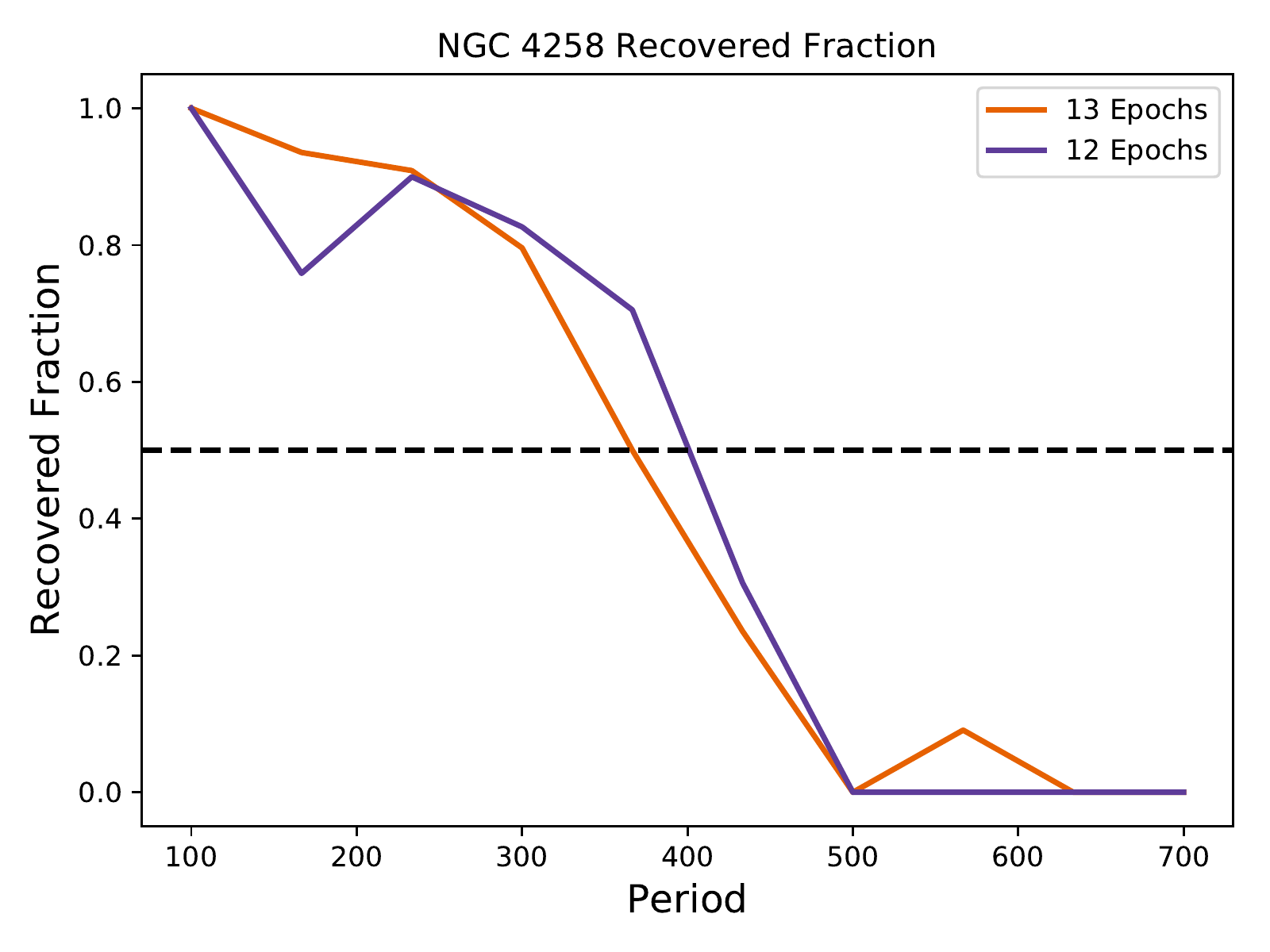}
  \caption{The fraction of Mira periods recovered as a function of period. The range of periods for the input Miras (from LMC Mira observations) ranged from 103 to 675 days. Miras were considered 'recovered' if their measured periods were within 30 days of their true periods. A dashed line is shown at a recovery fraction of 0.5 to guide the eye. For the vast majority of our sources, with 13 epochs of observation, the recovery rate $\sim 90 \%$.}
  \label{fig:recovery}
\end{figure}

We used a two-step method to measure the periods of the \ssample \, Mira-like objects remaining after the previous cuts based on variability and preliminary $\Delta \emph{F160W}$ amplitude. Unlike the previous steps, which used only the twelve recent epochs of \emph{F160W} data, we incorporated the additional epochs of observation from 2009 and early 2010 into our analysis at the next stage to get more accurate periods. 

First we calculated a Lomb-Scargle periodogram for each variable source. Then the peaks of the Lomb-Scargle were each fit with a Fourier series up to the third order. Well-sampled Mira light curves from the LMC have shown that Miras can have higher order harmonics but more could not be fit due to the limited number of available observations. We folded the light curves by each potential period and then fit every folded light curve using a Fourier series. We used a Bayesian information criterion (BIC) to determine if adding another harmonic to the fit was significant before increasing the number of harmonics. The BIC is defined as
\begin{equation}
\text{BIC} = -2\ln \mathcal{L} + k \ln(n)
\end{equation}
where $\mathcal{L}$ is the maximized value of the likelihood function for the estimated model, $n$ is the sample size, and $k$ is the number of free parameters to be estimated. We assumed that the errors were Gaussian and uncorrelated, and thus $\mathcal{L}$ was equal to the error variance of the fit with $k$ parameters.

Given the limited number of epochs, the BIC indicated that a simple sine function is most appropriate for almost all of the Miras. Finally, we used the Fourier fit parameters and period as initial guesses for a fit to the data using Levenberg-Marquardt least-squares curve-fitting. At this point all of the parameters were fit simultaneously. 

To verify that the periods were correct, we visually inspected each $P < 350$ day Mira candidate light curve and checked its fit to a sinusoid using the period determined earlier. Any periods that did not produce a good fit to the data were either refit by removing outliers and overriding the original fit, or, if a good fit could not be found, flagged as a lower-quality object and not used in the analysis. 

Since the baseline of our observation was only \baseline days, we only have at least one cycle of good phase coverage for Mira candidates with periods less than \baseline days. To calculate the period recovery rate, we used LMC Mira observations as templates and added photometric uncertainties and the observation sampling that reflected the NGC 4258 dataset. We tested both the case of 13 epochs of observation (incorporating in the earlier epoch from 2009) and 12 epochs of observation. We then measured the periods of our sample Miras and considered every period measured to within 30 days of the true input period as recovered. For both we found that the recovery rate of Mira periods less than 300 days was approximately 90$\%$. Figure \ref{fig:recovery} shows the results of the simulation.

\subsection{Samples and Selection Criteria}
\label{sec:samp}

\setlength{\tabcolsep}{.7em}
\begin{deluxetable*}{lccc}
\tabletypesize{\scriptsize}
\tablecaption{Mira Sample Criteria}
\tablewidth{0pt}
\tablehead{ & \colhead{Bronze} & \colhead{Silver} & \colhead{Gold}
}

\startdata

    Period Cut: & $P < 300$ days &  $P < 300$ days &   $P < 300$ days  \\
    Amplitude Cut: &$0.4 mag < \Delta \emph{F160W} < 0.8$ mag & $0.4 mag < \Delta \emph{F160W} < 0.8$ mag & $0.4 mag < \Delta \emph{F160W} < 0.8$ mag \\
    Color Cut: &$m_{F125W} - m_{F160W} < 1.3$ & $m_{F125W} - m_{F160W} < 1.3$ & $m_{F125W} - m_{F160W} < 1.3$ \\
    \emph{F814W} Detection: &-- & \emph{F814W} detection & Slope-fit to \emph{F814W} data$ > 3\sigma$  \\
    \emph{F814W} Amplitude: &-- & -- & $\Delta\emph{F814W} > 0.3$ mag \\

\enddata
\label{tab:samples}
\end{deluxetable*}

Our goal is to recover samples of the most secure Miras, rather than the most Miras, since the statistical uncertainty on the zeropoint of a PL due to our sample size will already be much smaller than the systematic error. We created three samples of Miras, which we have called Gold, Silver, and Bronze based on varying degrees of confidence in classification. Each sample contains predominantly O-rich Miras but the Bronze sample relies only on NIR information for classification, while the Gold and Silver samples are further vetted using our short time series of optical data. This makes the Bronze sample a good test case for future NIR-only observations of Miras. The criteria for each sample are given in Table \ref{tab:samples}. We outline the consequences and motivation for each of these criteria in the following sections.

\subsubsection{Period Cut}
\label{sec:percut}

\begin{figure}
\epsscale{1.2}
  \plotone{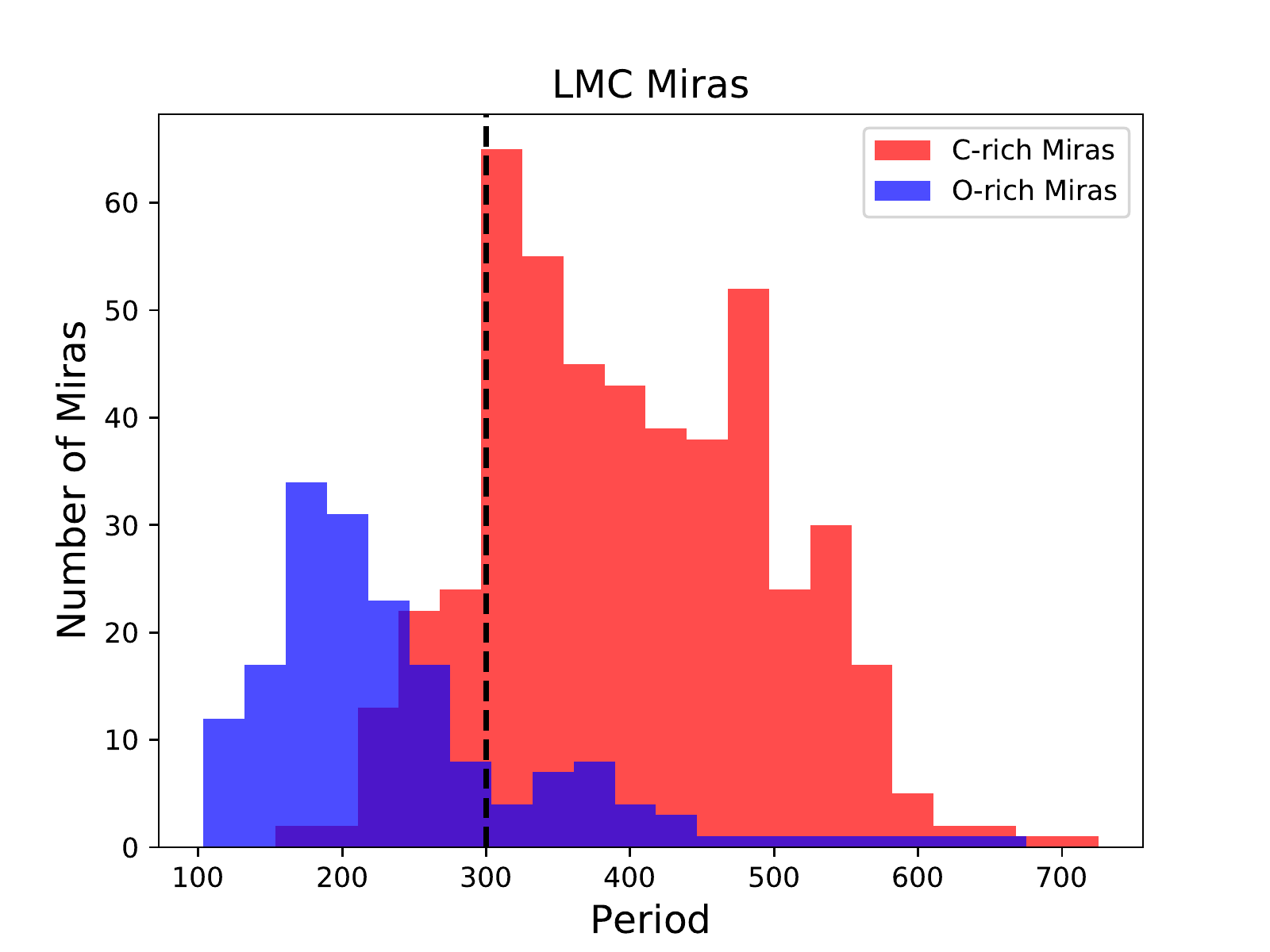}
  \caption{The distribution of periods for LMC O- and C-rich Mira variables. C-rich Mira periods are shown in red, the O-rich Mira periods are shown in blue. The black dashed line represents the period cut of 300 days.}
  \label{fig:perioddist}
\end{figure}

Because the baseline of our observations spanned only 305 days, we kept only Miras with periods less than 300 days. We also tested shorter upper period limits (down to 250 days) and found that there was no effect on the zeropoint. Choosing a more restrictive upper limit for the period cut will also make our sample less representative of the objects we would normally find in SNe Ia host galaxies. NGC 4258 is relatively close compared to most supernova hosts, so we will have an easier time finding sources at the longer end of the period range. A 300-day Mira is $\sim$ 0.5 mag more luminous than a 250-day Mira and as a result allows us to look in volume twice as large.

In addition to avoiding contamination from periods over 300 days that may be unreliable, limiting our sample to shorter-period Miras also has additional benefits for classification. The period cut reduces the number of C-rich stars in our sample since C-rich Miras typically have longer periods than O-rich Miras as discussed in \S \ref{sec:periods} (see Figure \ref{fig:allvariableshamp}). A distribution of C- and O-rich Mira periods in the LMC is shown in Figure \ref{fig:perioddist}. 

While this period cut will exclude longer-period O-rich Miras, some of these also do not make good distance indicators because they can be hot-bottom-burning stars (HBB). The onset of HBB depends on both mass and metallicity, but it is typically thought to occur in stars with initial masses greater than 4-5 $M_{\odot}$ that are near the end of their AGB phases \citep{Glass03}. \citet{Whitelock03} showed that HBB Miras deviate from a linear PLR, making them poor candidates for distance indicators. 

\subsubsection{F160W Amplitude}
\label{sec:hamp}

\begin{figure*}
\epsscale{1.2}
  \plotone{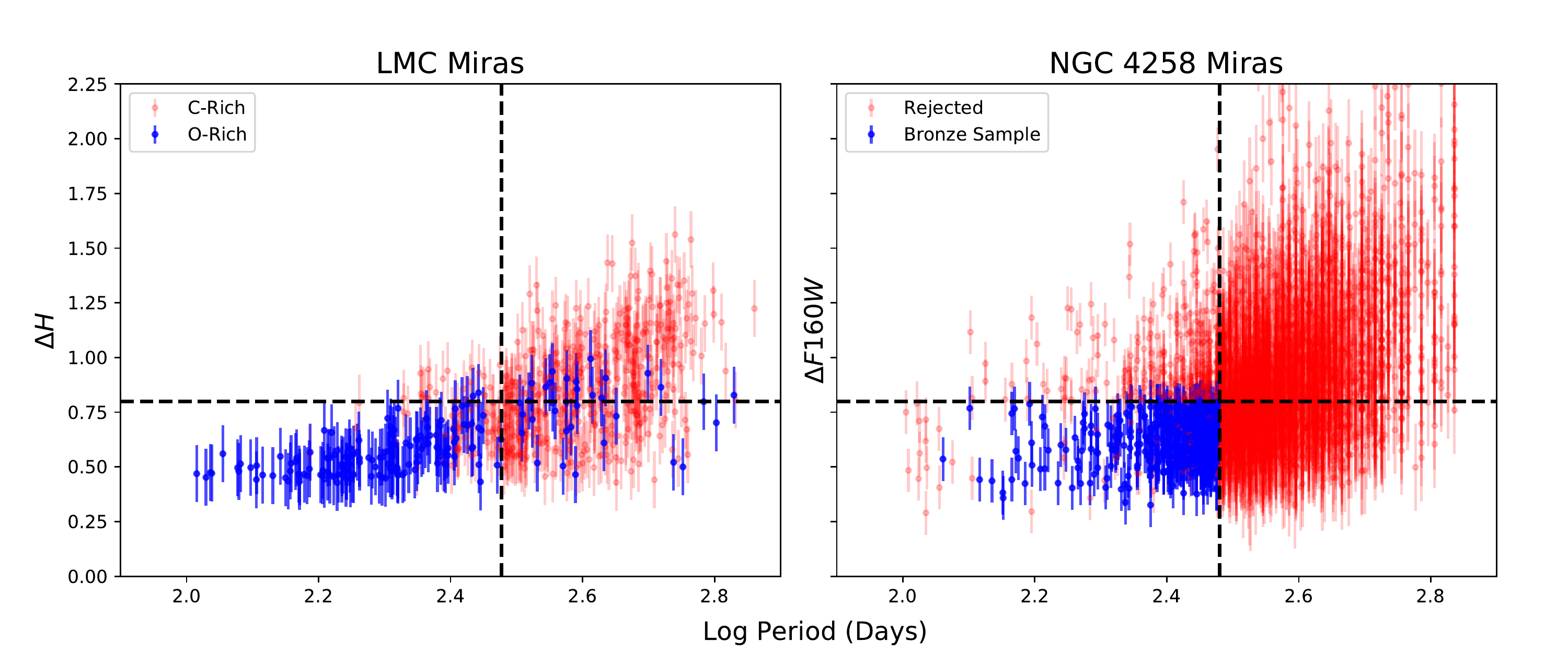}
  \caption{{\bf Left}: The amplitude and period relationship for LMC Miras. Red points are C-rich Miras and blue points are O-rich Miras as classified by \citet{Soszynski09b}. They divided the Mira sample along the $W_I$ \emph{vs} $W_{JK}$ plane, where $W_I$ and $W_{JK}$ are Wesenheit indices using optical and NIR magnitudes, respectively. Amplitude information of the LMC Miras comes from \citep{Yuan17b} with errors on the amplitude estimated to be $\sim$ 0.13 mag. The dashed vertical and horizontal lines represent the maximum period (300 days) and amplitude (0.8 mag) cuts respectively. Under our selection criterion, only objects within the bottom-left quadrant would have made it into the final Mira sample. {\bf Right}: The same plot constructed for NGC 4258 Miras with the objects in the Bronze Sample in blue, and all the rejected Miras in red. Period and amplitude beyond 300 days are unreliable but have been plotted with their best estimates shown for comparison. Errors on amplitudes greater than 300 days are estimated to be $\sim .2$ magnitudes. Periods below 300 days were verified by visual inspection.}
  \label{fig:ampvp}
\end{figure*}

Miras have generally been selected on the basis of their large optical ($V$ and $I$-band) amplitudes to distinguish them from semi-regular variables. Semi-regular variables (SRVs) can be as consistent as Miras in their variability and have similar periods but have smaller amplitudes than Miras. SRVs can also fall on the same PLR as Miras or on various other parallel PL relations \citep{Wood99, Trabucchi17}, depending on their pulsation mode. In general, SRVs are brighter than Miras with the same period and thus can bias the PLR if they are not removed from the final Mira sample. Previous studies of Miras \citep{Matsunaga09, Whitelock08} have suggested a minimum peak-to-trough variation of $\Delta J, \, \Delta H, \, \Delta K \sim 0.4$ mag to classify a variable as a Mira. Thus, we have used $\Delta \emph{F160W} > 0.4$ as the cutoff for minimum change in brightness over one cycle. 

In addition to removing SRVs, this minimum amplitude cut also allows us to remove constant stars and blended objects, which would not follow a PLR at all. For a variable star like a Mira or a Cepheid, the resulting blend will have a different color and amplitude from the original star in addition to being more luminous. 

O- and C-rich Miras can also have different amplitude distributions. \citet{Cioni03b} found that C-rich Miras had larger optical amplitudes on average than O-rich Miras in the SMC. \citet{Yuan17b} found that this was also the case for O-rich Miras in the LMC, especially when considering the amplitude distribution over many cycles. Over the course of a single cycle, C-rich Miras usually have larger amplitudes. This is caused by C-rich Miras having longer periods and thicker dust shells on average compared to O-rich Miras. Both longer-period Miras and heavily reddened stars are more likely to have larger amplitudes.
While there is considerable overlap in the distribution of amplitudes for O- and C-rich Miras (as shown in Figure \ref{fig:ampvp}) the largest-amplitude objects are usually C-rich.

The left half of Figure \ref{fig:ampvp} shows the distribution of LMC Miras as a function of period and $\Delta H$ using data from \citet{Yuan17b}. These Miras were classified by the OGLE team using a $W_I-W_{JK}$ diagram, where $W_I$ and $W_{JK}$ are the Wesenheit indicies in the optical and NIR, respectively \citep{Madore82}. Both types of Miras have estimated uncertainties in amplitude of $\sim 0.13$ mag but different distributions in amplitude.  Since we are interested in obtaining a clean sample of O-rich Miras, we employ a cut of $\Delta F160W < 0.8$ mag as a maximum amplitude cut in addition to the $ > 0.4$ mag minimum amplitude. The corresponding plot of accepted and rejected Miras in NGC 4258 is shown in the right half of Figure \ref{fig:ampvp}, with our `Bronze' Miras (Miras that met all of the NIR criteria) shown in blue. Some objects in the same quadrant as the Bronze sample were rejected on the basis of their uncertain periods. However, we expect that there should be a few percent overall contamination from C-rich Miras in this quadrant in the Gold sample. 

In the LMC, the ratio of C-to-O Miras in this quadrant is 1/3, and the overall C-to-O ratio in the LMC is 3/1. The C-to-O ratio in the solar neighborhood, which is more similar environment to the inner field of NGC 4258 is $\sim 1$ \citep{Ishihara11}. This suggests a $\sim 10 \%$ contamination rate from C-rich Miras from using these two cuts alone. Combining these with the optical observations and the color cut which both exclude the reddest objects, we estimate that the contamination in the Gold sample after all of the cuts should be a few percent. 

\subsubsection{Color}
\label{sec:jhcol}

\begin{figure}
\epsscale{1.2}
  \plotone{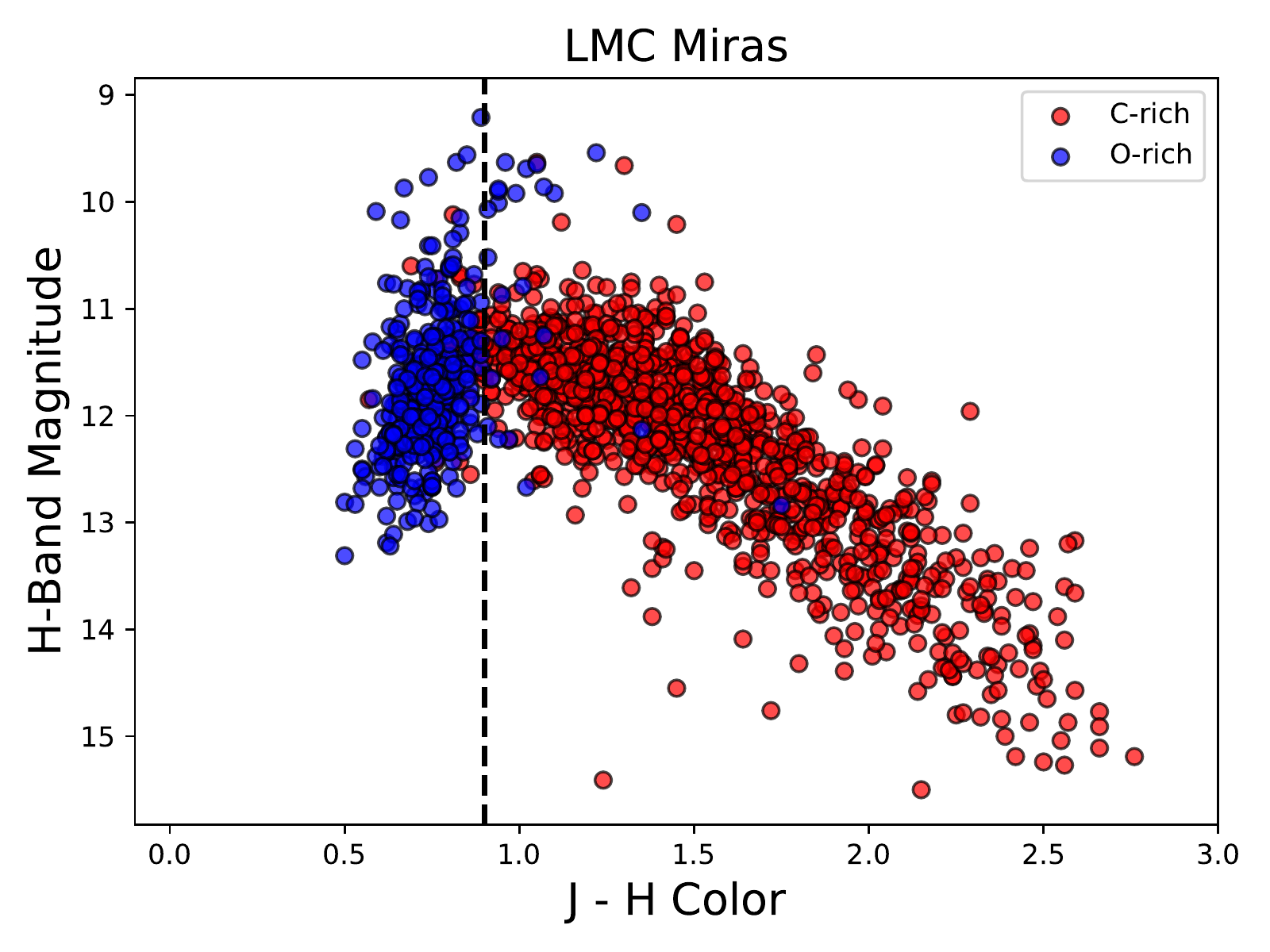}
  \plotone{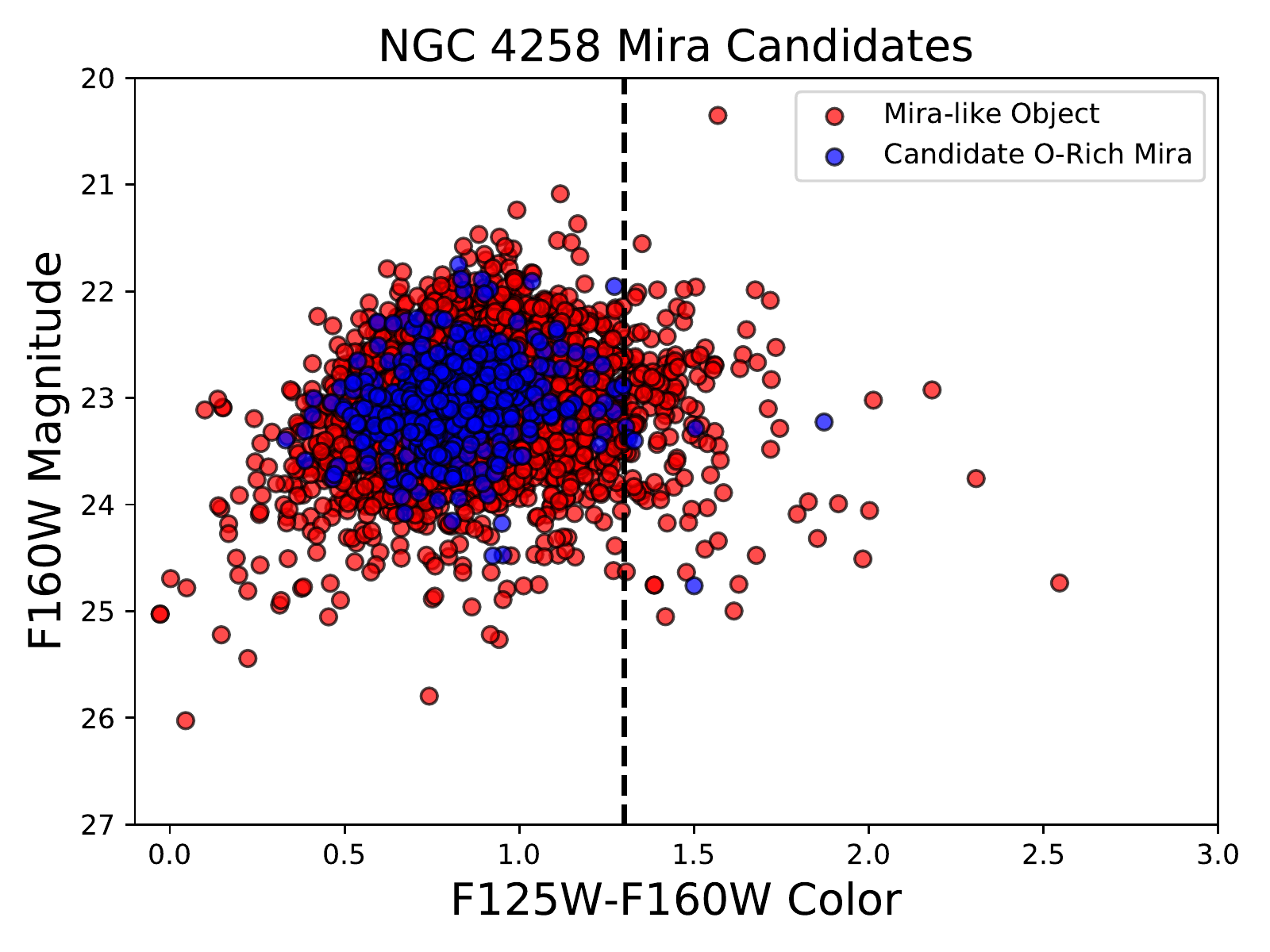}
  \plotone{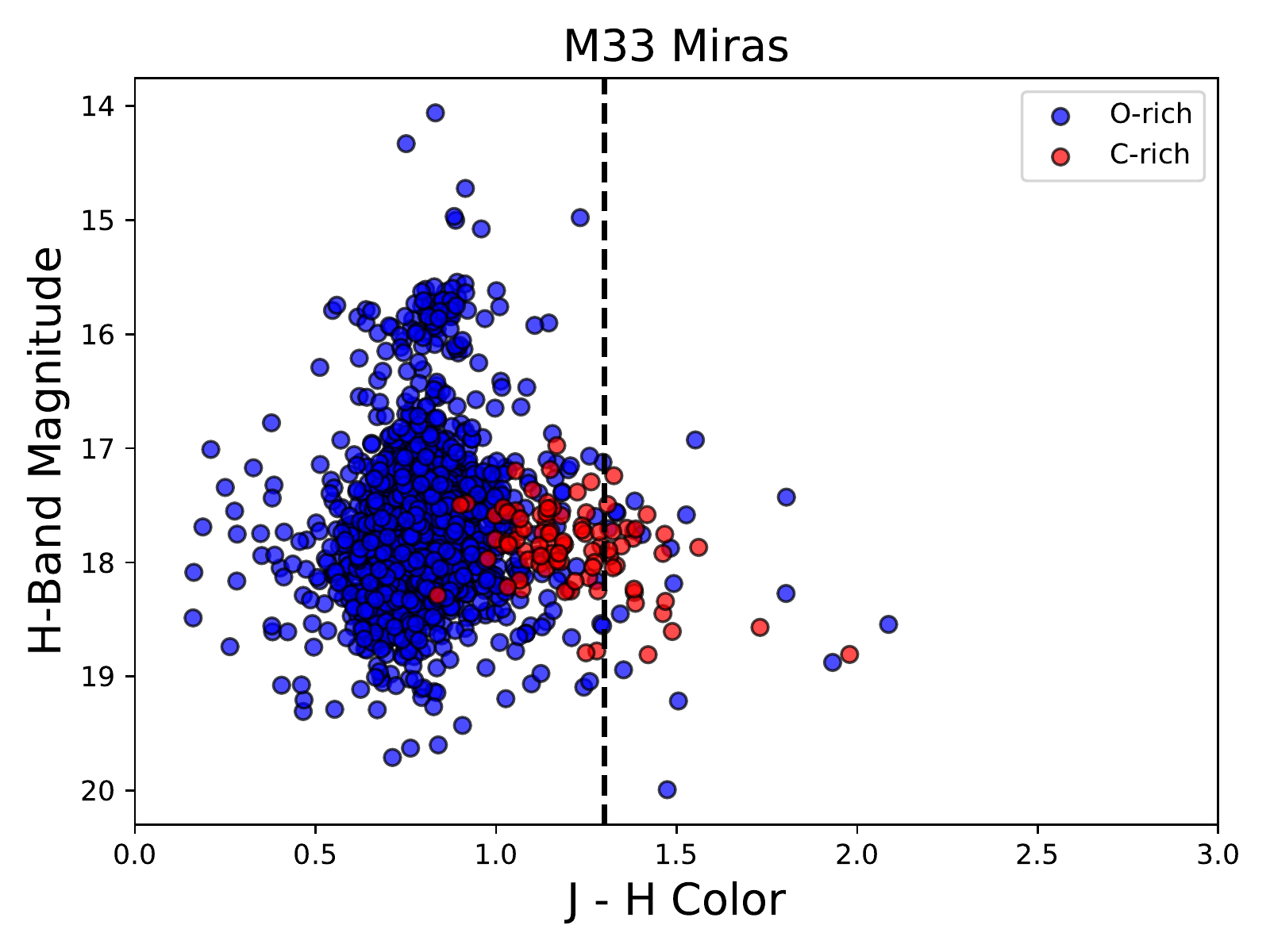}
  \caption{{\bf Top}: The color-magnitude diagram for Miras in the LMC made from cross-matching the OGLE-III star catalog with near-infrared data from IRSF.  The two subclasses of Miras were classified using a $W_I-W_{JK}$ diagram. The black dashed line is at $J-H = 0.9$.
  {\bf Middle}: The color-magnitude diagram for Mira candidates in the NGC 4258 made from comparing the mean \emph{F125W} and \emph{F160W} colors. The black dashed line represents $J-H = 1.3$. There are many more bluer variables in the NGC 4258 dataset than in the LMC data set. The `Candidate O-rich Miras' are objects that have passed all of our NIR cuts except for color.
  {\bf Bottom}: The color-magnitude diagram for Miras in M33, using UKIRT data from Yuan et al., in preparation. The black dashed line is at $J-H = 1.1$. These Miras were observed from the ground and detected in the optical. C-rich Miras, which are redder, as less likely to have been detected.  
  }
  \label{fig:cmds}
\end{figure}

We calculated the $\emph{F125W} - \emph{F160W}$ color by creating a stacked ``master" image in each bandpass where each epoch was weighted evenly (despite the first epoch in both bands having a longer observation time).  The $\emph{F125W} - \emph{F160W}$ color was then measured from the fluxes of the objects in these two images.

Our final NIR cut uses $\emph{F125W} - \emph{F160W}$ color. We see that the majority of C-rich Miras can be removed by employing a color cut of $J-H < 0.9$, as shown in Figure \ref{fig:cmds} (adopted from \citet{Soszynski09b}). Color cuts to separate O- and C-rich Miras are physically motivated by the differing opacity in the near-infrared and mid-infrared (MIR) in C- and O-rich stars. It is most effective to use both NIR and MIR colors for distinguishing between the two groups; broad NIR bands alone have not been shown to be sufficient to separate C- and O-rich Miras \citep{LeBertre94}. Medium-band filters on \emph{HST} that target unique spectral features have been used to separate C- and O-rich stars in M31 \citep{Boyer13, Boyer17}, but are not efficient for identifying Miras in more distant galaxies because they would require much longer integration times. 

The $J$ and $H$ filters are also better suited for distinguishing between C- and O-rich Miras than \emph{F125W} and \emph{F160W}, which are much more similar in their transmission functions. As seen in Figure \ref{fig:cmds}, the objects flagged as Mira-like (all objects both red and blue points) in our analysis do not follow the same distribution as the Miras in the LMC. The Mira-like objects in NGC 4258 appear to be one population rather than two seen in the LMC. This is most likely caused by differences in the \emph{HST} filter system and the typical ground-based NIR filters. Stellar models of C- and O-rich AGB stars from \citet{Aringer09} and \citet{Aringer16} suggest that these two populations overlap more in \emph{F125W}-\emph{F160W} color than in $J$ and $H$ color. Some of the reddest objects may also have been undetected. 

To avoid cutting through the middle of our population, we used $\emph{F125W}-\emph{F160W} < 1.3$ mag as a color cut to remove the reddest objects, but anticipated that it would not fully remove C-rich Miras from our sample. We employ the \emph{F160W} amplitude cuts from \S \ref{sec:hamp} and period cut to remove the majority of C-rich Miras instead. 

\begin{figure}
\epsscale{1.2}
  \plotone{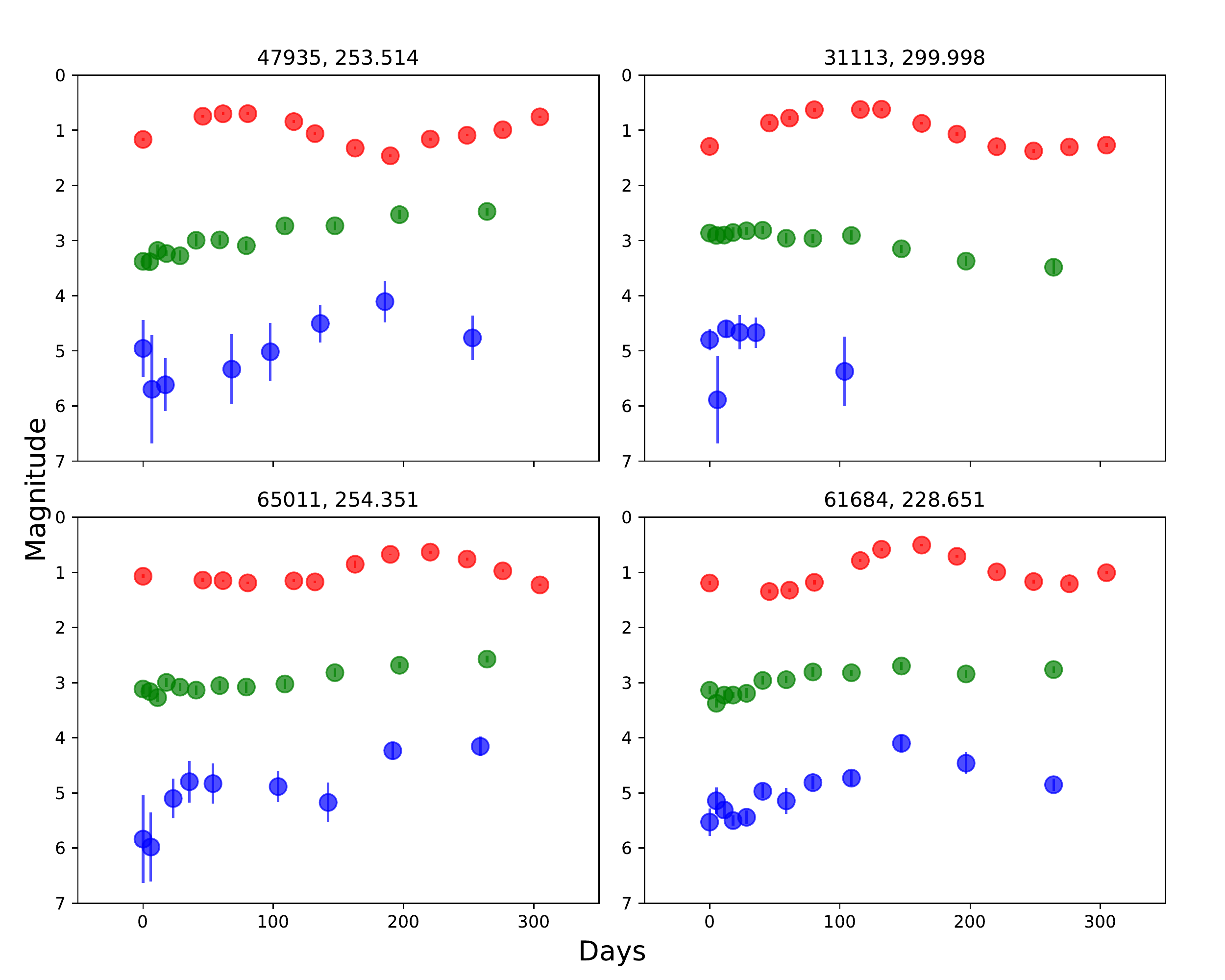}
  \caption{The \emph{F160W} (red), \emph{F814W} (green), and \emph{F555W} (blue) light curves for four Mira candidates. The magnitudes have been shifted in order to display every candidate Mira's light curve on the same plot axes and show the amplitude of the light curves. The horizontal axis marks the number of days since the start of each series of observations. The days for the optical light curves have been multiplied by a factor of six in order to better show the shape of the light curves. Numbers at the top of each subplot are photometry ID and calculated period of the object. As can be seen in Table \ref{tab:obs}, we did not have concurrent observations of near-infrared and optical data. Only 35 of the candidate Miras in the Bronze sample had \emph{F555W} light curves. \emph{F555W} data was not included in the analysis, but have been shown here for completeness.} 
  \label{fig:vihlcurves}
\end{figure}

\begin{figure}
\epsscale{1.2}
  \plotone{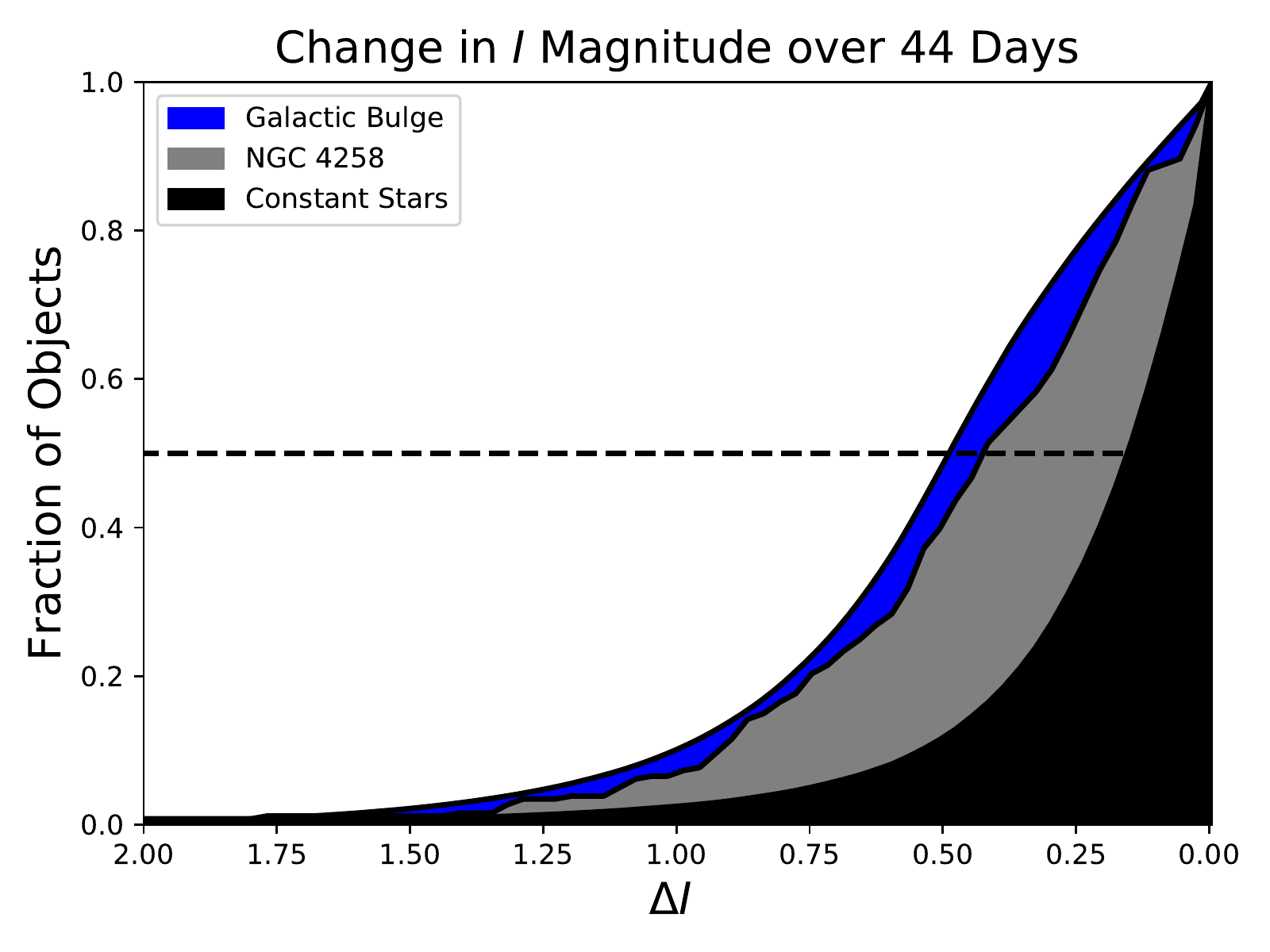}
  \caption{The change in $I$-band magnitude for Miras in the Galactic Bulge (detected with OGLE), Miras in the NGC 4258 Silver sample ($\silversize \,$ objects), and constant stars. The three curves show the cumulative distribution for each class of objects as a function of change in $I$-band magnitude over 44 days (the baseline of our \emph{F814W} observations.}
  \label{fig:cumulhist}
\end{figure}

\setlength{\tabcolsep}{.3em}
\begin{deluxetable}{lcc}
\tabletypesize{\scriptsize}
\tablecaption{Typical Mira Amplitudes and Absolute Magnitudes}
\tablewidth{0pt}
\tablehead{\colhead{Bandpass} & \colhead{Amplitude Range} & \colhead{Absolute Magnitude} \\
\colhead{} & \colhead{(mag)} & \colhead{(200-Day Mira)}}

\startdata

$V$ & 2.5- $>10$ & -1.4 \\
$I$ & 0.8-3.5 & -4.1 \\ 
$J$ & 0.5-3.0 & -5.8 \\
$H$ & 0.4-3.0 & -6.6 \\
$K_s$ & 0.4-3.0 & -7.0 \\

\enddata
\tablecomments{The range of amplitudes for Miras and the absolute magnitude of a 200-day Mira in various photometric bands. Upper limits on the amplitudes are approximate. $V$ band absolute magnitudes are not well known and have been estimated from OGLE data \citep{Soszynski09b}. The other absolute magnitudes are calculated using LMC and M33 PLRs from \citet{Yuan17a}.
}
\label{tab:magnitudes}
\end{deluxetable}

\subsubsection{F555W and F814W amplitudes}
\label{sec:viamp}

In addition to variability, amplitude, and color cuts based on NIR data which were applied to all three samples, we looked for corroborating evidence for the variability of Mira-like objects in \emph{F555W} and \emph{F814W} observations for the Silver and Gold samples. Given photospheric temperatures of only $\sim 3000-3500$ K, Miras are significantly less luminous at optical wavelengths. They also experience extremely large-amplitude variations in those bands, with $\Delta V \sim 10$ mag or greater in some cases. Table \ref{tab:magnitudes} shows a range of amplitudes in different bandpasses and magnitudes for a 200-day Mira. 

We estimated the recovery rate of Miras in the optical data. Using the \emph{F160W} observations, we calculated the \emph{F160W} phase at the time of the optical observations and assumed a phase lag between the optical and NIR phases that was dependent on period. The phase lag was calculated using data from \citet{Yuan17b}. We used
\begin{equation}
\phi_I - \phi_H = 2\pi(0.469 - 0.144\log(P))
\end{equation}
where $\phi_I$ is the $I$-band phase, $\phi_H$ is the $H$-band phase, and $P$ is the period. We assumed that the $I$ and $H$ bands are roughly equivalent in phase to the \emph{F814W} and \emph{F160W}, respectively. We used the estimated differences in mean magnitude and amplitude from Table \ref{tab:magnitudes} to convert from \emph{F160W} magnitudes and amplitudes to \emph{F555W} and \emph{F814W}. Any stars that had a signal-to-noise ratio greater than 3 were considered recovered in the simulation. This resulted in an expected recovery rate of $\sim 78\%$ for \emph{F814W}. 

We used the DAOMATCH and DAOMASTER programs to match the NIR and optical master source lists and found that 296 out of 438 Miras (or $68\%$) of the Miras from the Bronze sample were matched with sources in \emph{F814W} images. This is roughly in agreement with the simulated expected numbers. As anticipated, very few of our Mira sample could be matched with \emph{F555W} light curves (a $3\%$ recovery rate was predicted by the simulation, compared to $2\%$ in reality). Thus, we used only information from the \emph{F814W} observations in our selection criteria.
Light curves for some of the few potential Miras with both \emph{F555W} and \emph{F814W} matches are shown alongside their \emph{F160W} light curves in Figure \ref{fig:vihlcurves}. The optical observations precede the NIR epochs by about ten years. 

We interpret a Mira's chances of \emph{F814W} detection as a function of both a Mira's $\emph{F814W}- \emph{F160W}$ color and phase. We assumed that all of the Miras in the simulation had the same color, but C-rich or heavily dust-enshrouded O-rich Miras are known to be very red and would be difficult to detect in optical bandpasses. Similarly, Miras measured towards the trough of their light cycles would be more difficult to detect than Miras at the peak. However, we were unable to determine our Mira candidates' true phases at the time of the previous observations, so we could not test this directly through simulations.   

In order to determine the significance of a change in magnitude over the observation baseline, we fit each \emph{F814W} light curve fragment with a linear fit. We then kept only objects with at least a naive `$3\sigma$' significance in change of magnitude in the Gold Sample. This allows us to check that the object we detected as variable in \emph{F160W} is variable in \emph{F814W}. 

We also used the difference in magnitude between the first and last epoch of observation to estimate the I-band amplitude. Thus, we could determine if the sources that were variable in \emph{F160W} were also variable in the optical observations. Given that the baseline of the optical observations were only 44 days and the shortest-period Miras have periods of about 100 days, the \emph{F555W} and \emph{F814W} light curves cover only about 15-20\% of an average Mira's oscillation and at most contain only about $\sim 60\%$ of its total variation. Due to the short temporal baseline of observations, objects close to the peak or trough (if detected) of their observations will be rejected on account of their small overall variations.

We used the OGLE Galactic-Bulge sample of Miras \citep{Soszynski13} to calculate the distribution of changes in $I$ band magnitude expected over a period of 44 days. We compared this with the distribution we obtained while looking at the changes in \emph{F814W} magnitude over the same duration baseline for objects in our Silver sample. However, the two bandpasses are not completely identical and their distributions are slightly different, as shown in Figure \ref{fig:cumulhist}. We also created some light curves of constant stars with photometric noise added in as a null case. The Silver sample from NGC 4258 and the Galactic Bulge distribution both show significantly higher levels of $I$ band variation than the simulated ``constant" stars over the same period, suggesting that they are true variable stars. The optical requirement also excludes the reddest stars, which are most likely C-rich.

\section{Systematics}
\label{sec:sys}

In order to get an accurate relative distance and to compare our results with previous studies of Miras using ground-based observations, we needed to transform ground-based $H$ and $J$ magnitudes of LMC Miras into the \emph{HST F160W} magnitude. We performed artificial star tests to account for excess background due to the density of sources in our field and loss of flux due to an imperfect PSF model. Each of these introduces a systematic error into the final result. 

To estimate the systematic errors of each of our cuts, we varied each cut around a standard deviation of the values chosen or the range of values present in the literature when possible and looked at the effect that it had on the zeropoint of the PLR relation. Table \ref{tab:sys} has a summary of each of these contributions.

\setlength{\tabcolsep}{.7em}
\begin{deluxetable*}{lcccrrccc}
\tabletypesize{\scriptsize}
\tablecaption{Final Sample of Miras}
\tablewidth{0pt}
\tablehead{\colhead{ID} & \colhead{Period} & \colhead{R.A.} & \colhead{Dec.} & \colhead{X} & \colhead{Y} & \colhead{Magnitude} & \colhead{Amplitude} & \colhead{Quality} \\
\colhead{} & \colhead{(Days)} & \colhead{(J2000.0)} & \colhead{(J2000.0)} & \colhead{(Pixels)} & \colhead{(Pixels)} & \colhead{(F160W mag)} & \colhead{($\Delta F160W$)}  & \colhead{}
}

\startdata

    60841 &  287.015 &   12 18 47.087   &     +47 19 32.85 & 1210.179 & 1249.445 &   22.029 &    0.776 &     Gold \\ 
    47935 &  253.514 &   12 18 42.138   &     +47 20 42.89 &  165.583 &  983.280 &   22.918 &    0.684 &     Gold \\ 
    44958 &  272.561 &   12 18 43.976   &     +47 20 33.73 &  418.105 &  920.864 &   22.790 &    0.712 &     Gold \\ 
    24235 &  292.075 &   12 18 44.806   &     +47 21 10.89 &  200.187 &  497.320 &   22.455 &    0.716 &     Gold \\ 
    31113 &  299.998 &   12 18 44.585   &     +47 20 58.19 &  280.781 &  636.955 &   22.539 &    0.796 &     Gold \\ 
    52981 &  275.948 &   12 18 42.945   &     +47 20 25.17 &  386.517 & 1087.064 &   22.523 &    0.746 &   Bronze \\ 
    64439 &  256.965 &   12 18 41.148   &     +47 20 15.63 &  288.277 & 1325.193 &   22.938 &    0.712 &     Gold \\ 
    64343 &  298.885 &   12 18 40.925   &     +47 20 17.73 &  249.748 & 1323.194 &   22.792 &    0.758 &   Silver \\ 
    33823 &  265.960 &   12 18 45.165   &     +47 20 47.61 &  422.330 &  690.870 &   22.681 &    0.654 &   Silver \\ 
    39445 &  237.358 &   12 18 42.893   &     +47 20 54.83 &  143.229 &  807.290 &   22.897 &    0.798 &     Gold \\

\enddata
\tablecomments{A partial list of Miras is shown here for information regarding form and content.} 
\label{tab:miratable}
\end{deluxetable*}

\setlength{\tabcolsep}{.3em}
\begin{deluxetable}{lc}
\tabletypesize{\scriptsize}
\tablecaption{Systematic Uncertainties}
\tablewidth{0pt}
\tablehead{\colhead{Systematic} & \colhead{Uncertainty} }

\startdata

\emph{F160W} Amplitude Cut................ & \amperr \\
\emph{F160W} - \emph{F125W} Color Cut .......\ & \colorcuterr \\ 
Color Correction Term ................. & \jhcolorerr \\
Intrinsic Scatter............................ & \scattererr \\
Total............................................. & \syserr \\

\enddata
\tablecomments{The approximate contribution to the total systematic error of the gold sample from each systematic.} 
\label{tab:sys}
\end{deluxetable}

\subsection{Slope}
\label{sec:sysslope}

Miras and other variable stars are typically fit with a linear Period-Luminosity Relation (called the Leavitt Law for Cepheids). However, there is some evidence for break at 10 days for Cepheids and previous Mira observations have suggested that the Mira PLR may have a break as well, at periods of about 400 days \citep{Ita11}. This is likely to be caused by the onset of HBB. 

\citet{Yuan17a} used a quadratic fit for the Period-Luminosity relation instead of fitting the sample with two linear PLRs with a break. We fit each of our subsamples of Miras using their quadratic PL relations for the $H$ band, which is the closest match to \emph{F160W}:
\begin{align} \label{eq:plr}
m = a_0 - 3.59(\log P - 2.3) -3.40(\log P - 2.3)^2 
\end{align}
where $P$ is the period in days, $m$ is the $H$-band magnitude, and $a_0$ is what we have called the zeropoint. We fit for $a_0$ and hold the other parameters fixed to the values from \citet{Yuan17a}. 

This PLR was derived using observations of about 170 LMC O-rich Miras. While the \emph{F160W} filter is bluer than the $H$ band, deriving a PLR from the NGC 4258 data alone was not possible because of the much larger scatter induced by the background brightness fluctuations, which is discussed in great detail in \S \ref{sec:crowd}. With observations of Miras in more galaxies using \emph{F160W}, we could simultaneously fit the Miras in multiple galaxies and derive a more robust fit in this bandpass.

\subsection{Artificial Star Tests}
\label{sec:crowd}

The high density of sources and unresolved background objects in our images will systematically bias our magnitude measurements towards brighter values. This is caused by the superposition of several point sources. We refer to this effect as crowding. 

We correct for crowding using artificial star tests. Starting with a master image created from combining all twelve epochs of \emph{F160W} data, we use DAOPHOT to place fake sources in the images at the same apparent magnitude as the Mira candidates. We then compare the recovered magnitudes of the artificial stars with the input magnitudes and adjust the input magnitudes to better agree with the recovered magnitudes. On average, the artificial stars were measured to be 0.25 magnitudes more luminous than their true magnitudes.

We then repeat the steps in the photometry process up to the creation of the master source list and then compare the recovered and input magnitudes to determine the crowding correction. Because there are $\sim 1700$ variables in the image, we add in one fake star for \emph{half} of the variable stars in the image at any given time to avoid artificially raising the background of the image. The artificial stars are dropped within a 25-pixel radius of each Mira, and at least 10 pixels away from the edges of the image. Only stars that did not fall within 3 pixels of another star up to 1.5 magnitudes fainter were used in the analysis. 

After calculating the difference in the input and recovered magnitudes, we then use a three sigma-clip about the median to remove outliers. The mean difference between input and recovered magnitude for each star is then the crowding correction we apply. 

\subsection{Mean Magnitude Correction}
\label{sec:meanmag} 

We used OGLE O-rich Miras cross-matched with $J$ and $H$ magnitudes from the Infrared Survey Facility (IRSF) catalogue \citep{Kato07} to determine the LMC zeropoint. 
The mean magnitudes for NGC 4258 were defined as the first term in the Fourier-series fit to each object. Because Mira light curves are irregular in shape, they do not spend the same amount of time at each phase in their cycles, creating a small bias in single-epoch measurements when compared to mean magnitudes.

We used Monte Carlo simulations to calculate the difference between these two estimates of mean magnitude and found that on average, the PLRs measured using the fit mean magnitudes were 0.02 fainter than the PLRs measured using single-epoch mean magnitudes. Therefore, in order to correct between the two, we added in a mean magnitude correction of -0.02 mag to our final results.

\section{Results}
\label{sec:results}

\subsection{Color Transformation}
\label{sec:calib}

\begin{figure}
\epsscale{1.2}
  \plotone{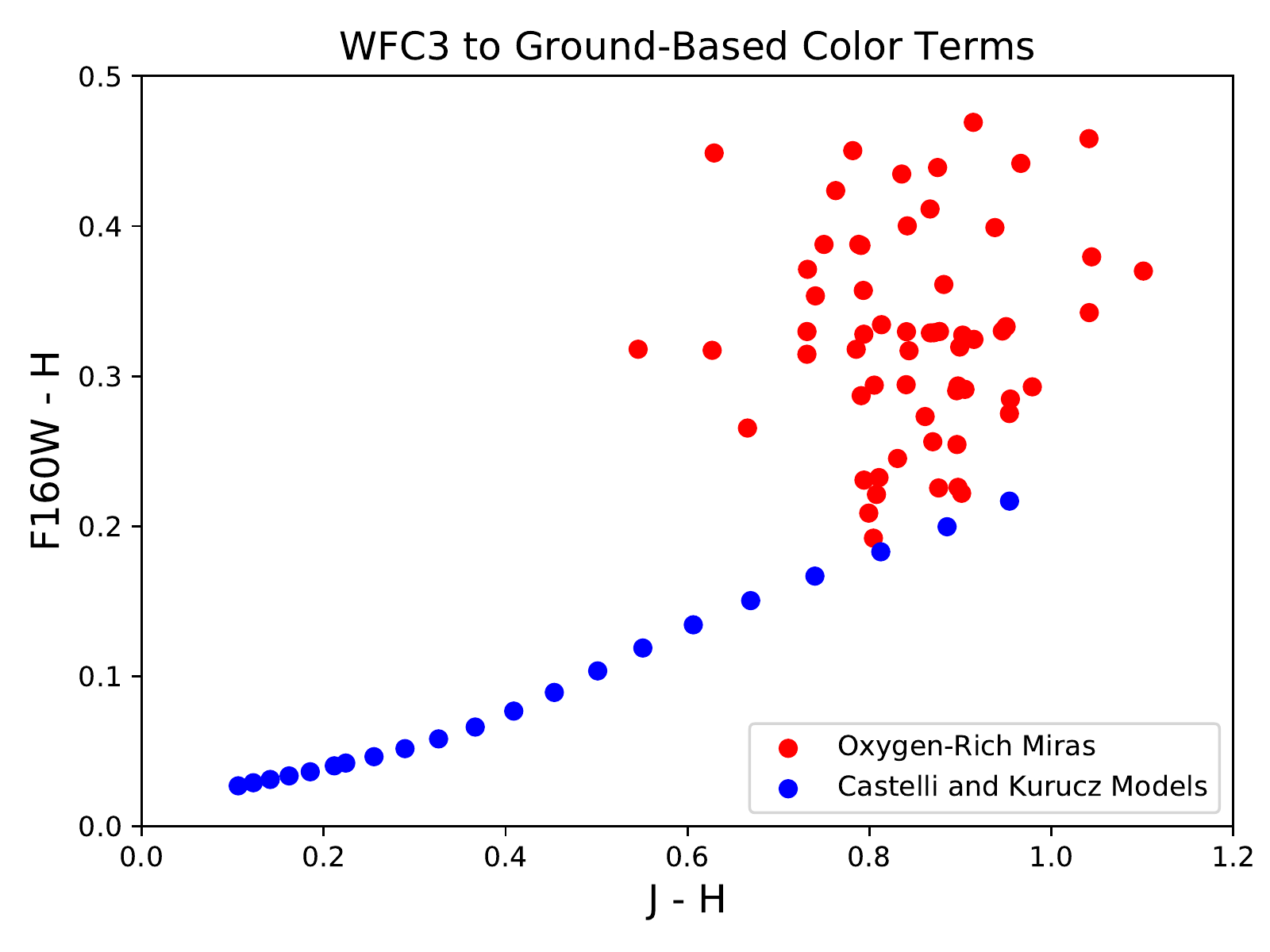}
  \caption{Non-linear color term for the \emph{F160W} - $H$ transformation as a function of $J-H$ color. The blue points represent calculations using \citet{Castelli04} models with $3500 < T < 7000$, $\log g=0.1$ and solar metallicity. The red points are based on observed spectra of O-rich Miras. Note that neither set of points has a constant color coefficient as a function of color. The mean of the red points is 0.39, which we adopt for our transformations.
  }
  \label{fig:colorterm}
\end{figure}

\begin{figure}
\epsscale{1.2}
  \plotone{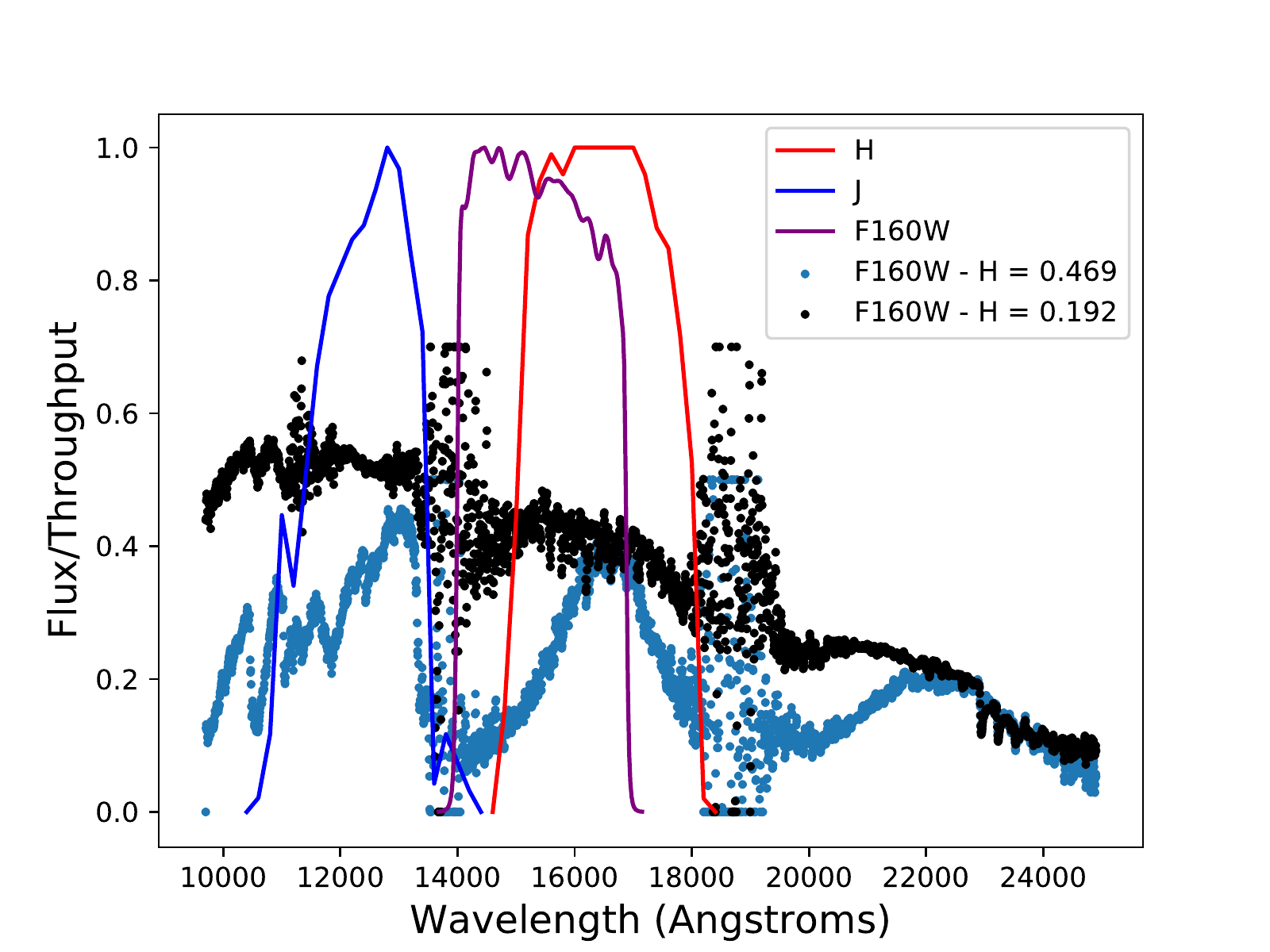}
  \caption{The reddest and bluest O-rich Mira spectra in used to calculate the color correction (very red OH/IR stars were excluded) are shown in blue and black respectively with ground-based $J$ and $H$ filters and \emph{HST} WFC3 \emph{F160W}. The filters have been normalized to have a peak throughput of 1. The difference in color appears to be the result of a combination of both the continuum emission and the much stronger absorption lines in the redder spectrum.}
  \label{fig:ospectra}
\end{figure}

\begin{figure}
\epsscale{1.2}
  \plotone{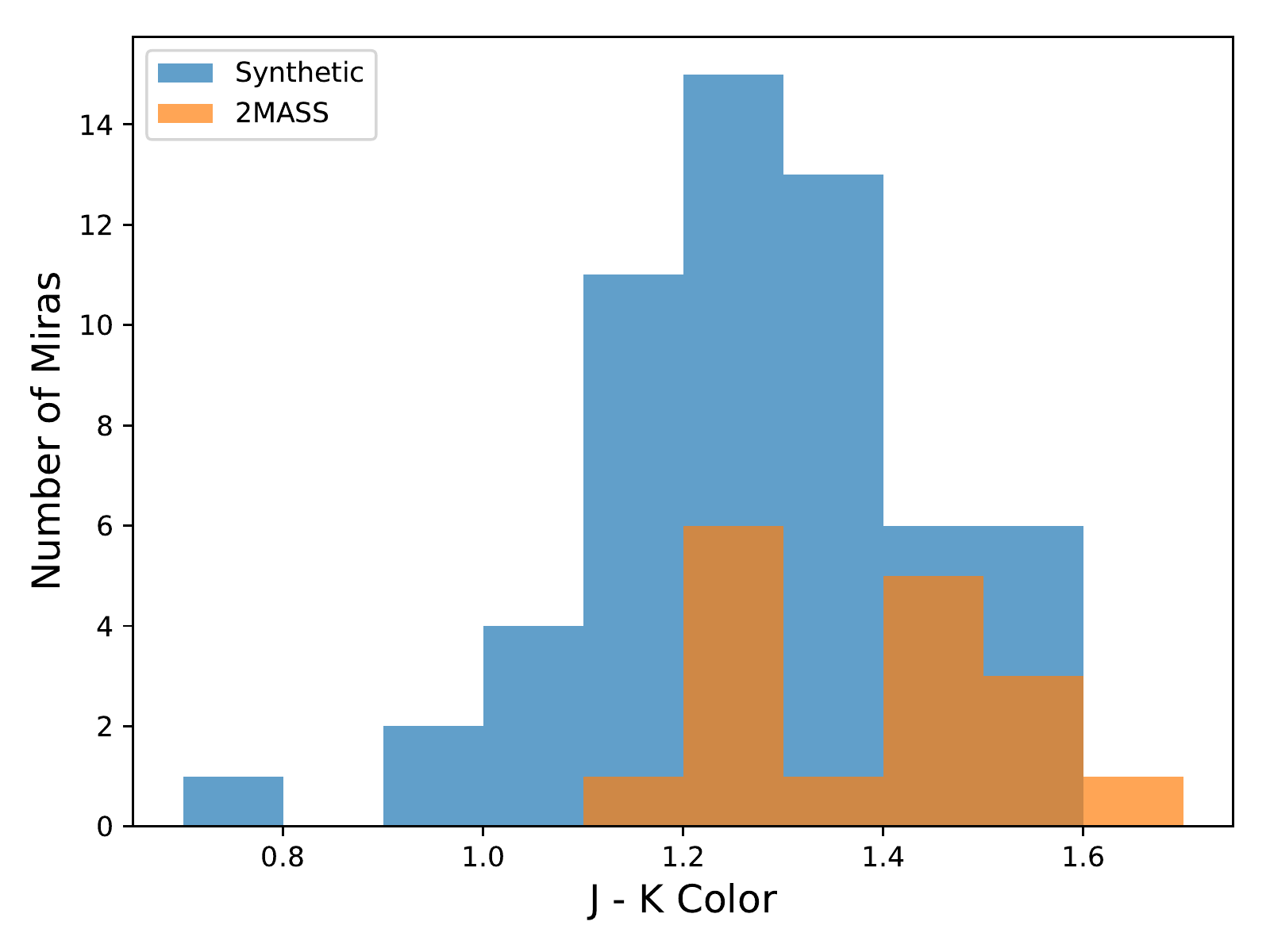}
  \caption{A comparison of the $J-K$ color derived synthetically using PySynPhot and $J-K$ colors from 2MASS. Most of the objects in the Galactic sample of Miras were observed at more than one part of their light cycle and thus had synthetic colors that varied by $\sim~0.15$ magnitudes. The synthetic colors had an overall standard deviation of 0.19 magnitudes whereas the 2MASS colors had a standard deviation of 0.14 magnitudes.}
  \label{fig:JKcol}
\end{figure}

In order to compare ground-based NIR PLRs with the \emph{F160W} PLR from NGC 4258, we calculated a color correction to transform the ground-based \emph{H} band Mira observations to \emph{F160W} observations. Miras have heavy molecular absorption lines in the NIR compared to M-type main sequence stars so we used real O-rich Mira NIR spectra observed from the ground from \citet{Lancon00} as input to PySynPhot to derive the transformations for Miras specifically \citep{PySynPhot}.

This resulted in an $H$ to \emph{F160W} transformation of:
\begin{align} \label{eq:transformation}
\emph{F160W} = H + 0.39(J-H)
\end{align}

This is a significantly larger color term than was found for the bluer Cepheids (0.16) in R16. This difference is due to the non-linearity of the color term, shown in Figure \ref{fig:colorterm}. Because Cepheids are much bluer than Miras (a typical Cepheid is an $F$ star), and the color term is non-linear, we must use a different color transformation of Miras. Using the same color term to transform from $H$ to \emph{F160W} for Miras as we used for Cepheids would result in a $\sim 0.18$ mag difference for an average O-rich Mira with a $J-H$ color of 0.8. The ground-based spectra are affected by the telluric absorption bands, as seen in Figure \ref{fig:ospectra}. Water bands dominate the near infrared spectra of O-rich Miras, and these are difficult to separate from telluric features in ground-based observations.  

Because the O-rich Miras we used to calculate the color transformation displayed a large scatter in \emph{F160W} - $H$ color, we also examined spectra of individual Miras that were particularly blue or red to check that the spectra were calibrated well enough for synthetic photometry. We found that very red Miras had much larger absorption features (the result of having more dust) than very blue Miras. Additionally, we compared the measured $J-K$ colors of the O-rich Miras in the 2MASS catalog to the $J-K$ colors derived synthetically (Figure \ref{fig:JKcol}). We found that for individual Miras, the two were in agreement and the standard deviation of the two distributions of color (0.19 from spectrophotometry and 0.14 from 2MASS measurements) was also similar. The 2MASS colors were on average slightly redder (by $\sim 0.1$ magnitudes), suggesting that the true color correction might be slightly larger. However, the distribution in Mira color appears to be real and not the result of poor calibration.

\subsection{Mira Samples}

\begin{figure*}
\epsscale{1.2}
  \plotone{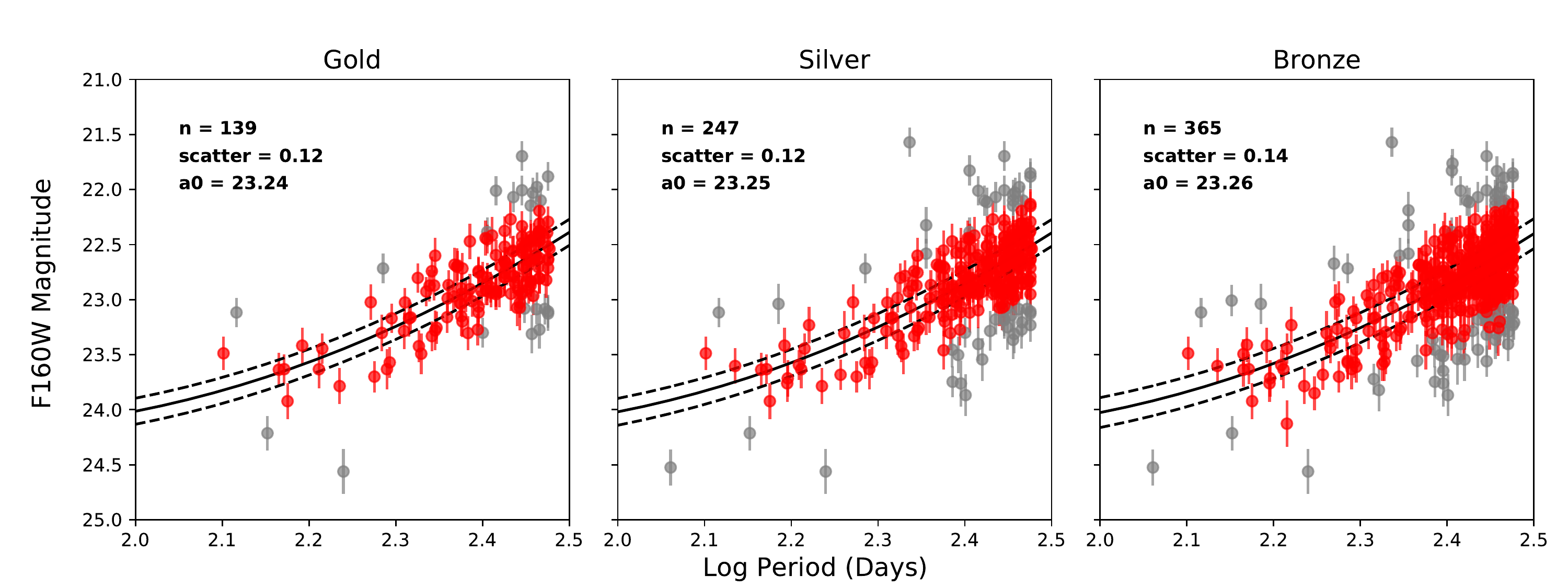}
  \caption{Mira Period-Luminosity relations for the Gold, Silver, and Bronze subsamples (left, center, and right, respectively). Red points denote objects used in the final fit, while gray points represent variables that were removed through iterative $3\sigma$ clipping. The solid black curves show the best-fit relations, while the dashed lines denote the $1\sigma$ scatter (0.11, 0.13, 0.14 $\sim$ mag respectively). The functional forms of the PLRs are from \citet{Yuan17a} and only the zeropoint was fit.}
  \label{fig:goldPL}
\end{figure*}

We created three subsets of Mira candidates based on our estimate of their reliability as distance indicators and the possibility of contamination, applying the cuts discussed described in Table \ref{tab:samples}. For our Gold sample of Mira candidates, we had a total final sample size of \totalgoldsize \, objects. Fitting these objects to the PL relation derived by \citet{Yuan17a} for the $H$ band gave us a zeropoint of $a_0 = \goldzpt \pm \goldrerr$ mag. The sigma clipping removed 22, $\sim 14 \%$ of the Gold sample Mira candidates from the final PLR, leaving a total of 139 Mira candidates remaining.

For the larger Silver sample, we had a final sample size of \totalsilversize \, and determined $a_0 = \silverzpt \pm \silverrerr$ mag for this sample. After sigma clipping, which removed 48 objects, $ \sim 16 \%$ of the Mira candidates, leaving 248 remaining for the fit.

Finally, the Bronze sample consisted of \totalbronzesize \, objects and had a zeropoint of $a_0 = \bronzezpt \pm \bronzererr$ mag. We sigma-clipped out 72 Mira candidates, $\sim 16 \% $, comparable to the amount removed in the Silver sample. The PLR for each sample is shown in Figure \ref{fig:goldPL}. 

Despite the different selection criteria, the zeropoints of all of the samples are almost the same and the fraction of their outliers is also consistent. This is especially important for the Bronze sample, which used only NIR criteria. It suggests that future studies can also be successful with only NIR \emph{HST} data to select Miras.

\subsection{Relative Distance to the LMC}

We compared our results in NGC 4258 with a sample of O-rich Miras discovered in the LMC by the OGLE survey \citep{Soszynski09b} to determine the relative distance modulus between these two galaxies. We obtained random-phase $J$ and $H$ magnitudes for the LMC variables from the IRSF catalog of \citet{Kato07}. We used Equation 3 to transform these into the equivalent \emph{F160W} magnitudes, which were fit with the PLR of Equation \ref{eq:plr} to solve for the zeropoint. The difference between the NGC 4258 and LMC zeropoints yields the relative distance modulus (not corrected for small differences in foreground extinction between the two galaxies). 

With the Gold sample defined in the previous section, we calculated a distance modulus relative to the LMC of $\Delta \mu_g = \goldmod \pm \goldrerr_r \pm \syserr_{sys} $ mag using Miras from the OGLE survey. For the Silver sample, we have $\Delta \mu_s = \silvermod \pm \silverrerr_r \pm \silversyserr_{sys} $ mag, and for the Bronze sample, $\Delta \mu_b = \bronzemod \pm \bronzererr_r \pm \bronzesyserr_{sys}$ mag. 
These are consistent with a previous measurement of the Cepheid relative distance modulus from R16, $\Delta \mu_{R16} = 10.92 \pm 0.02$ mag. 

In order for the Cepheid scale to agree with the Planck results, it would need to be too short by $\sim 0.20$ mag. Currently the Cepheids and Miras give consistent relative distances, but we can consider a hypothetical Mira relative distance to NGC 4258 of $10.75 \pm 0.07$ mag, in agreement with Planck. We find that it disagrees with the Cepheid relative distance from R16 by $2.4 \sigma$.  This demonstrates that both the Cepheid and Mira distance scales have some tension with the Planck results. Because the Cepheid and Mira results are independent, we also would not expect these discrepancies with Planck to be caused by the same effect. 

Finally, we also used the color transformation from \S \ref{sec:calib} to derive the PLR coefficients for an \emph{F160W}-band PLR using the ground $J$ and $H$-band relations from \citet{Yuan17b}. We then refit the PLRs for for both the LMC and NGC 4258 and found that these two methods yield marginal differences in the results. 

\subsection{Absolute Calibration to NGC 4258}

We use the improved megamaser distance to NGC 4258 from R16 of $29.387 \pm 0.057$ mag. The uncertainty in the \citet{Humphreys13} value was reduced to 2.6\% from 3\% by increasing the number of Monte Carlo Markov Chain (MCMC) trial values in the analysis by a factor of a hundred. Using this distance modulus puts the absolute calibration of the PLR for the Gold sample at $a_0 = \goldabsmag$ mag in the \emph{F160W} bandpass. 

\subsection{Spatial Distribution}
\label{sec:spatialdist}

\begin{figure*}
\epsscale{1.2}
  \plotone{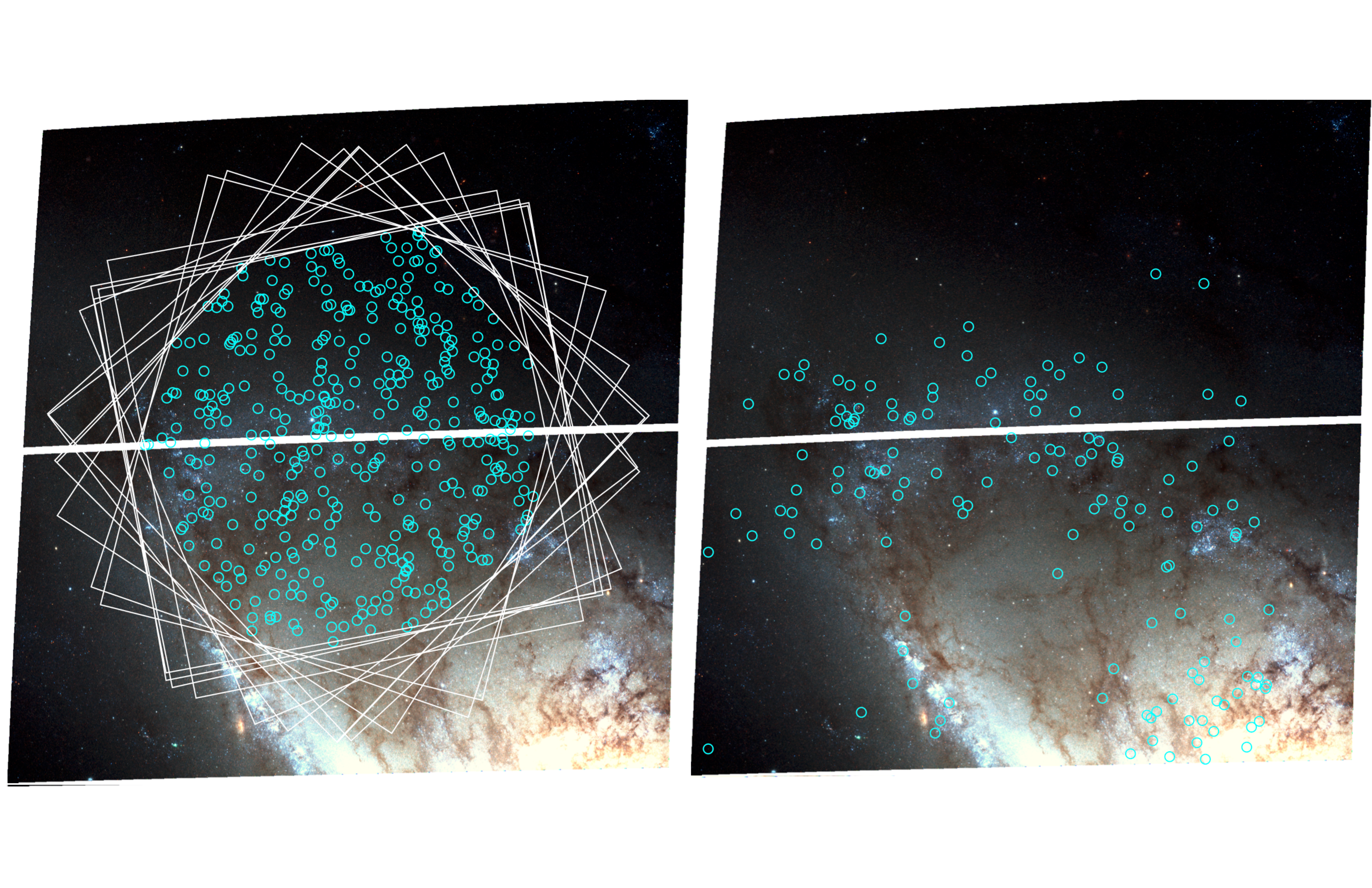}
  \caption{{\bf Left:} The locations of the Bronze sample Miras (cyan circles) in in the NGC 4258 ACS inner field. The white regions show the \emph{F160W} footprint. Because our observations were taken over the course of one year, the orientation changed in each observation, leaving an approximately circular area that was observed in all epochs. We searched in this region only for Miras. {\bf Right: }Cepheids (cyan circles) from \citet{Macri06} on top of the same ACS field. The Cepheid distribution traces the spiral arms of the galaxy while the Miras are more common and can be found evenly across the smaller \emph{F160W} footprint.}
  \label{fig:mirasreg}
\end{figure*}

\begin{figure}
\epsscale{1.2}
  \plotone{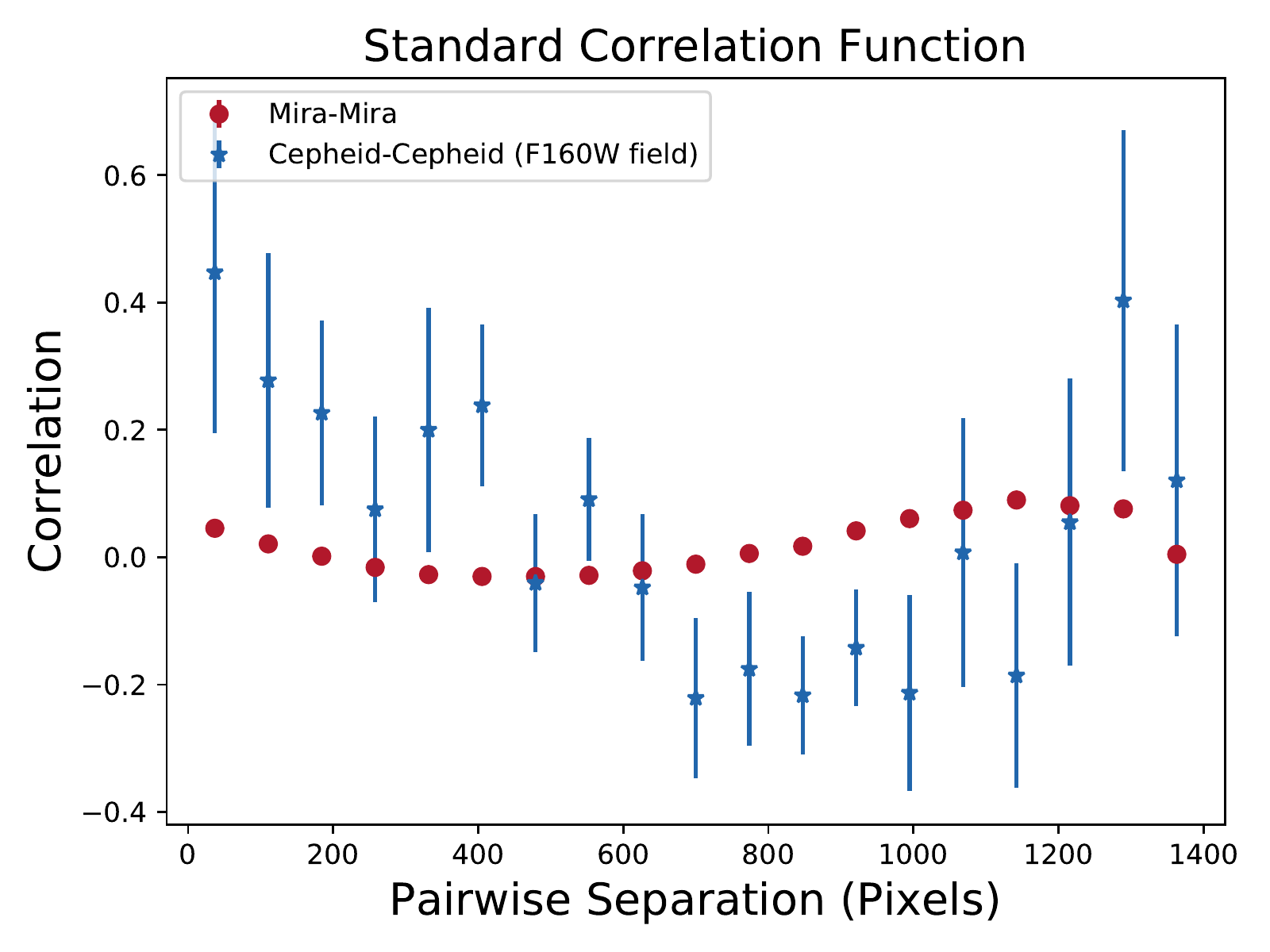}
  \caption{The Cepheid-Cepheid and Mira-Mira standard correlation function. in the NGC 4258 \emph{F160W} field. Errors are obtained through bootstrap resampling. Cepheids are blue, the Miras in red. Due to the small sample of Cepheids (84) compared to Miras (\totalbronzesize), the errors for the Cepheid-Cepheid autocorrelation functions are much larger.}
  \label{fig:correlation}
\end{figure}

As a sanity check, we compared the spatial distributions of our Bronze Mira candidate sample with the spatial distribution of Cepheids in NGC 4258. Figure \ref{fig:mirasreg} shows the locations of Cepheids and Miras in the galaxy overlaid with the \emph{F160W} footprints. The Cepheids trace the spiral arm of the galaxy while the Miras are found randomly distributed in the \emph{F160W} footprint. These differences in spatial distribution have a physical origin in the progenitors of Cepheids and Miras. Cepheids, with their intermediate and high-mass progenitors, are young stars that are only found in regions with active star formation. Thus, they are present in the denser spiral arms of a galaxy and are part of the disk population. Almost all Miras have progenitors of low-to-intermediate mass, which therefore have intermediate-to-old ages and can exist in areas without recent star formation. For our sample limited to short-period Miras only, this is especially true, since the progenitor stars will all be of low mass. Miras can additionally be found in both the disk and halo populations. 

We calculated the autocorrelation functions for both Cepheids and Miras in the \emph{F160W} footprint shown in Figure \ref{fig:mirasreg}. The autocorrelation function for Miras discovered in this project and the Cepheids discovered by \citet{Macri06} is shown in Figure \ref{fig:correlation}. The Cepheid and Mira autocorrelation functions follow different distributions, with the Mira autocorrelation being much flatter, as expected for evenly distributed objects. The results confirm that the two distributions are different spatially, and agree with what we would expect from a physical understanding of Mira and Cepheid progenitors. 

\section{Discussion}
\label{sec:discussion}

The largest source of uncertainty in local measurements of $H_0$ remains the number of SNe Ia host galaxies that have been calibrated with Cepheid distances. A 300-day Mira is roughly comparable in \emph{F160W} brightness to a 30-day Cepheid, allowing them to be observed to approximately the same volume. SNe Ia used in R16 all have modern photometry, low reddening ($A_V < 0.5$ mag), and observations prior to peak luminosity. Additionally, only late-type host galaxies were targeted to ensure the presence of Cepheids. Using Miras over Cepheids would increase the number of SN host galaxies for cross-calibration and eliminate potential biases caused by host galaxy morphology. Recent papers such as \citet{Jones15} and \citet{Rigault15} have disagreed on whether host galaxy morphology can have an effect on the luminositites of Type Ia SNe. \citet{Rigault15} found that Type Ia SNe in locally star-forming environments are dimmer than SNe Ia in locally passive environments. \citet{Jones15} also searched for this effect but found that there was little evidence for a difference. Regardless of the effect local environment can have we can avoid this potential problem altogether by using Miras.

Since Miras are an older population star they can be found in most galaxies regardless of host galaxy morphology. This can help us create a sample of cross-calibrators that is more representative of the Hubble flow SNe Ia sample. \citet{Rejkuba04} was able to derive a K-band Mira PLR for the giant elliptical galaxy NGC 5128, which would have been an unlikely target as an SNe Ia calibrator host. Miras are also part of the halo population so we can also potentially look for Miras in hosts that are not face-on, further increasing the number of potential targets. 

\setlength{\tabcolsep}{.7em}
\begin{deluxetable*}{lcccrrccc}
\tabletypesize{\scriptsize}
\tablecaption{LMC and SMC Miras}
\tablewidth{0pt}
\tablehead{\colhead{Period} & \colhead{O-rich SMC} & \colhead{C-rich SMC} & \colhead{O-rich LMC} & \colhead{C-rich LMC}} \\

\startdata

    $\leq 300$ days & 12 & 80 & 349 & 187 \\ 
    $\leq 250$ days & 7 & 18 & 302 & 17 \\ 

\enddata
\tablecomments{The numbers of O- and C-rich Miras in the SMC \citep{Soszynski11} and LMC \citep{Soszynski09b} from the OGLE survey.} 
\label{tab:mirastatistics}
\end{deluxetable*}

Below 300 days, the proportion of O-to-C-rich Miras increases as period decreases. The host galaxy's metallicity and initial mass function will also affect the relative proportions of O- to C-rich Miras, but most Miras with periods less than 300 days will be O-rich up to about Fe/H $\sim -1.0$.  In the Small Magellanic Cloud (SMC), with a Fe/H $\sim - 1.0$ , even the short-period ranges ($< 250$ days ) are dominated by C-rich stars, as shown in Table \ref{tab:mirastatistics}. We would be able to find more O-rich Miras and other variables that fall on the sample PLR by searching for objects with periods below 100 days. At longer wavelengths C-rich stars can serve as distance indicators as well. \citet{Whitelock17} compares the O-rich and C-rich Mira PLRs in $K_s$. This would not work for \emph{HST}, since there are no filters redder than \emph{F160W} but may be possible for \emph{JWST}.

In the future we will be able to further reduce our uncertainties in the relative distance by directly comparing measurements of Miras observed in the \emph{HST} NIR with our calibrated PLRs from NGC 4258. In addition, by using the same criteria and filters we can help ensure that we selected for the same classes of objects. The consistency of the results between the Bronze, Silver, and Gold samples also demonstrates that Miras can we conduct this search without optical data.

\section{Conclusions}
\label{sec:conc}

1. We discovered \totalbronzesize \, Mira candidates in one field of NGC 4258 using only the \emph{HST} \emph{F160W} bandpass.

2. We developed criteria to reduce contamination from C-rich Miras in our sample that do not use spectroscopic information and found that amplitude cuts can help separate C- and O-rich Miras. This allows us to discover and characterize Mira candidates using solely or primarily \emph{F160W}. 

3. We determined a relative distance modulus between NGC 4258 and the Large Magellanic Cloud based on Mira variables of $\Delta \mu_g = \goldmod \pm \goldrerr_r \pm \syserr_{sys}$ mag, which is consistent with the Cepheid relative distance modulus and also consistent with the relative distance modulus obtained using geometric methods. 

4. We have calibrated the extragalactic Mira distance ladder in \emph{F160W} using the geometric distance to NGC 4258.  

5. We derived a Mira-specific color transformation from the ground-based $H$-band to the \emph{HST} \emph{F160W}. 

\section{Acknowledgments}
\label{sec:ack}

This project was funded through \emph{HST} grants associated with the following programs: GO-13445, GO-11570, GO-9810. We acknowledge with thanks the $o$Ceti observations from the AAVSO International Database contributed by observers worldwide and used in this research. PAW acknowledges a research grant from the South African National Research Foundation. We would also like to thank Drs. Yoshifusa Ita, Noriyuki Matsunaga, Gregory Sloan, Martha Boyer, Steven McDonald, Daniel Shafer, and Sjoert van Velzen for their helpful discussions.

\bibliographystyle{apj}
\bibliography{n4258}

\begin{thebibliography}{}
\expandafter\ifx\csname natexlab\endcsname\relax\def\natexlab#1{#1}\fi

\bibitem[{{Addison} {et~al.}(2017){Addison}, {Watts}, {Bennett}, {Halpern},
  {Hinshaw}, \& {Weiland}}]{Addison17}
{Addison}, G.~E., {Watts}, D.~J., {Bennett}, C.~L., {et~al.} 2017, ArXiv
  e-prints, arXiv:1707.06547

\bibitem[{{Aringer} {et~al.}(2016){Aringer}, {Girardi}, {Nowotny}, {Marigo}, \&
  {Bressan}}]{Aringer16}
{Aringer}, B., {Girardi}, L., {Nowotny}, W., {Marigo}, P., \& {Bressan}, A.
  2016, \mnras, 457, 3611

\bibitem[{{Aringer} {et~al.}(2009){Aringer}, {Girardi}, {Nowotny}, {Marigo}, \&
  {Lederer}}]{Aringer09}
{Aringer}, B., {Girardi}, L., {Nowotny}, W., {Marigo}, P., \& {Lederer}, M.~T.
  2009, \aap, 503, 913

\bibitem[{{Bhardwaj} {et~al.}(2017){Bhardwaj}, {Macri}, {Rejkuba}, {Kanbur},
  {Ngeow}, \& {Singh}}]{Bhardwaj17}
{Bhardwaj}, A., {Macri}, L.~M., {Rejkuba}, M., {et~al.} 2017, \aj, 153, 154

\bibitem[{{Blanco} \& {McCarthy}(1983)}]{Blanco83}
{Blanco}, V.~M., \& {McCarthy}, M.~F. 1983, \aj, 88, 1442

\bibitem[{{Bono} {et~al.}(2010){Bono}, {Caputo}, {Marconi}, \&
  {Musella}}]{Bono10}
{Bono}, G., {Caputo}, F., {Marconi}, M., \& {Musella}, I. 2010, \apj, 715, 277

\bibitem[{{Bonvin} {et~al.}(2017){Bonvin}, {Courbin}, {Suyu}, {Marshall},
  {Rusu}, {Sluse}, {Tewes}, {Wong}, {Collett}, {Fassnacht}, {Treu}, {Auger},
  {Hilbert}, {Koopmans}, {Meylan}, {Rumbaugh}, {Sonnenfeld}, \&
  {Spiniello}}]{Bonvin17}
{Bonvin}, V., {Courbin}, F., {Suyu}, S.~H., {et~al.} 2017, \mnras, 465, 4914

\bibitem[{{Boyer} {et~al.}(2013){Boyer}, {Girardi}, {Marigo}, {Williams},
  {Aringer}, {Nowotny}, {Rosenfield}, {Dorman}, {Guhathakurta}, {Dalcanton},
  {Melbourne}, {Olsen}, \& {Weisz}}]{Boyer13}
{Boyer}, M.~L., {Girardi}, L., {Marigo}, P., {et~al.} 2013, \apj, 774, 83

\bibitem[{{Boyer} {et~al.}(2017){Boyer}, {McQuinn}, {Groenewegen}, {Zijlstra},
  {Whitelock}, {van Loon}, {Sonneborn}, {Sloan}, {Skillman}, {Meixner},
  {McDonald}, {Jones}, {Javadi}, {Gehrz}, {Britavskiy}, \& {Bonanos}}]{Boyer17}
{Boyer}, M.~L., {McQuinn}, K.~B.~W., {Groenewegen}, M.~A.~T., {et~al.} 2017,
  ArXiv e-prints, arXiv:1711.02129

\bibitem[{{Bresolin}(2011)}]{Bresolin11}
{Bresolin}, F. 2011, \apj, 729, 56

\bibitem[{{Castelli} \& {Kurucz}(2004)}]{Castelli04}
{Castelli}, F., \& {Kurucz}, R.~L. 2004, ArXiv Astrophysics e-prints,
  astro-ph/0405087

\bibitem[{{Cioni} \& {Habing}(2003)}]{Cioni03a}
{Cioni}, M.-R.~L., \& {Habing}, H.~J. 2003, \aap, 402, 133

\bibitem[{{Cioni} {et~al.}(2001){Cioni}, {Marquette}, {Loup}, {Azzopardi},
  {Habing}, {Lasserre}, \& {Lesquoy}}]{Cioni01}
{Cioni}, M.-R.~L., {Marquette}, J.-B., {Loup}, C., {et~al.} 2001, \aap, 377,
  945

\bibitem[{{Cioni} {et~al.}(2003){Cioni}, {Blommaert}, {Groenewegen}, {Habing},
  {Hron}, {Kerschbaum}, {Loup}, {Omont}, {van Loon}, {Whitelock}, \&
  {Zijlstra}}]{Cioni03b}
{Cioni}, M.-R.~L., {Blommaert}, J.~A.~D.~L., {Groenewegen}, M.~A.~T., {et~al.}
  2003, \aap, 406, 51

\bibitem[{{Dhawan} {et~al.}(2017){Dhawan}, {Jha}, \& {Leibundgut}}]{Dhawan17}
{Dhawan}, S., {Jha}, S.~W., \& {Leibundgut}, B. 2017, ArXiv e-prints,
  arXiv:1707.00715

\bibitem[{{Follin} \& {Knox}(2017)}]{Follin17}
{Follin}, B., \& {Knox}, L. 2017, ArXiv e-prints, arXiv:1707.01175

\bibitem[{{Ford} {et~al.}(2003){Ford}, {Clampin}, {Hartig}, {Illingworth},
  {Sirianni}, {Martel}, {Meurer}, {McCann}, {Sullivan}, {Bartko}, {Benitez},
  {Blakeslee}, {Bouwens}, {Broadhurst}, {Brown}, {Burrows}, {Campbell},
  {Cheng}, {Feldman}, {Franx}, {Golimowski}, {Gronwall}, {Kimble}, {Krist},
  {Lesser}, {Magee}, {Miley}, {Postman}, {Rafal}, {Rosati}, {Sparks}, {Tran},
  {Tsvetanov}, {Volmer}, {White}, \& {Woodruff}}]{Ford03}
{Ford}, H.~C., {Clampin}, M., {Hartig}, G.~F., {et~al.} 2003, in \procspie,
  Vol. 4854, Future EUV/UV and Visible Space Astrophysics Missions and
  Instrumentation., ed. J.~C. {Blades} \& O.~H.~W. {Siegmund}, 81--94

\bibitem[{{Fraser} {et~al.}(2008){Fraser}, {Hawley}, \& {Cook}}]{Fraser08}
{Fraser}, O.~J., {Hawley}, S.~L., \& {Cook}, K.~H. 2008, \aj, 136, 1242

\bibitem[{{Freedman} {et~al.}(2001){Freedman}, {Madore}, {Gibson}, {Ferrarese},
  {Kelson}, {Sakai}, {Mould}, {Kennicutt}, {Ford}, {Graham}, {Huchra},
  {Hughes}, {Illingworth}, {Macri}, \& {Stetson}}]{Freedman01}
{Freedman}, W.~L., {Madore}, B.~F., {Gibson}, B.~K., {et~al.} 2001, \apj, 553,
  47

\bibitem[{{Glass} \& {Lloyd Evans}(2003)}]{Glass03}
{Glass}, I.~S., \& {Lloyd Evans}, T. 2003, \mnras, 343, 67

\bibitem[{{Gonzaga} {et~al.}(2012){Gonzaga}, {Hack}, {Fruchter}, \&
  {Mack}}]{Gonzaga12}
{Gonzaga}, S., {Hack}, W., {Fruchter}, A., \& {Mack}, J. 2012, {The DrizzlePac
  Handbook}

\bibitem[{{Hamren} {et~al.}(2015){Hamren}, {Rockosi}, {Guhathakurta}, {Boyer},
  {Smith}, {Dalcanton}, {Gregersen}, {Seth}, {Lewis}, {Williams}, {Toloba},
  {Girardi}, {Dorman}, {Gilbert}, \& {Weisz}}]{Hamren15}
{Hamren}, K.~M., {Rockosi}, C.~M., {Guhathakurta}, P., {et~al.} 2015, \apj,
  810, 60

\bibitem[{{Houk}(1963)}]{Houk63}
{Houk}, N. 1963, \aj, 68, 253

\bibitem[{{Humphreys} {et~al.}(2013){Humphreys}, {Reid}, {Moran}, {Greenhill},
  \& {Argon}}]{Humphreys13}
{Humphreys}, E.~M.~L., {Reid}, M.~J., {Moran}, J.~M., {Greenhill}, L.~J., \&
  {Argon}, A.~L. 2013, \apj, 775, 13

\bibitem[{{Iben} \& {Renzini}(1983)}]{Iben83}
{Iben}, Jr., I., \& {Renzini}, A. 1983, \araa, 21, 271

\bibitem[{{Inno} {et~al.}(2015){Inno}, {Matsunaga}, {Romaniello}, {Bono},
  {Monson}, {Ferraro}, {Iannicola}, {Persson}, {Buonanno}, {Freedman},
  {Gieren}, {Groenewegen}, {Ita}, {Laney}, {Lemasle}, {Madore}, {Nagayama},
  {Nakada}, {Nonino}, {Pietrzy{\'n}ski}, {Primas}, {Scowcroft},
  {Soszy{\'n}ski}, {Tanab{\'e}}, \& {Udalski}}]{Inno15}
{Inno}, L., {Matsunaga}, N., {Romaniello}, M., {et~al.} 2015, \aap, 576, A30

\bibitem[{{Ishihara} {et~al.}(2011){Ishihara}, {Kaneda}, {Onaka}, {Ita},
  {Matsuura}, \& {Matsunaga}}]{Ishihara11}
{Ishihara}, D., {Kaneda}, H., {Onaka}, T., {et~al.} 2011, \aap, 534, A79

\bibitem[{{Ita} \& {Matsunaga}(2011)}]{Ita11}
{Ita}, Y., \& {Matsunaga}, N. 2011, \mnras, 412, 2345

\bibitem[{{Ita} {et~al.}(2004){Ita}, {Tanab{\'e}}, {Matsunaga}, {Nakajima},
  {Nagashima}, {Nagayama}, {Kato}, {Kurita}, {Nagata}, {Sato}, {Tamura},
  {Nakaya}, \& {Nakada}}]{Ita04}
{Ita}, Y., {Tanab{\'e}}, T., {Matsunaga}, N., {et~al.} 2004, \mnras, 347, 720

\bibitem[{{Jang} \& {Lee}(2017)}]{Jang17}
{Jang}, I.~S., \& {Lee}, M.~G. 2017, \apj, 835, 28

\bibitem[{{Jones} {et~al.}(2015){Jones}, {Riess}, \& {Scolnic}}]{Jones15}
{Jones}, D.~O., {Riess}, A.~G., \& {Scolnic}, D.~M. 2015, \apj, 812, 31

\bibitem[{{Jones} {et~al.}(1996){Jones}, {Carney}, \& {Fulbright}}]{Jones96}
{Jones}, R.~V., {Carney}, B.~W., \& {Fulbright}, J.~P. 1996, \pasp, 108, 877

\bibitem[{{Kato} {et~al.}(2007){Kato}, {Nagashima}, {Nagayama}, {Kurita},
  {Koerwer}, {Kawai}, {Yamamuro}, {Zenno}, {Nishiyama}, {Baba}, {Kadowaki},
  {Haba}, {Hatano}, {Shimizu}, {Nishimura}, {Nagata}, {Sato}, {Murai},
  {Kawazu}, {Nakajima}, {Nakaya}, {Kandori}, {Kusakabe}, {Ishihara},
  {Kaneyasu}, {Hashimoto}, {Tamura}, {Tanab{\'e}}, {Ita}, {Matsunaga},
  {Nakada}, {Sugitani}, {Wakamatsu}, {Glass}, {Feast}, {Menzies}, {Whitelock},
  {Fourie}, {Stoffels}, {Evans}, \& {Hasegawa}}]{Kato07}
{Kato}, D., {Nagashima}, C., {Nagayama}, T., {et~al.} 2007, \pasj, 59, 615

\bibitem[{{Kholopov} {et~al.}(1985){Kholopov}, {Samus}, {Kazarovets}, \&
  {Perova}}]{Kholopov85}
{Kholopov}, P.~N., {Samus}, N.~N., {Kazarovets}, E.~V., \& {Perova}, N.~B.
  1985, Information Bulletin on Variable Stars, 2681

\bibitem[{{Lan{\c c}on} \& {Wood}(2000)}]{Lancon00}
{Lan{\c c}on}, A., \& {Wood}, P.~R. 2000, \aaps, 146, 217

\bibitem[{{Le Bertre} {et~al.}(1994){Le Bertre}, {Epchtein}, {Guglielmo}, \&
  {Le Sinadier}}]{LeBertre94}
{Le Bertre}, T., {Epchtein}, N., {Guglielmo}, F., \& {Le Sinadier}, P. 1994,
  \apss, 217, 105

\bibitem[{{Macri} {et~al.}(2015){Macri}, {Ngeow}, {Kanbur}, {Mahzooni}, \&
  {Smitka}}]{Macri15}
{Macri}, L.~M., {Ngeow}, C.-C., {Kanbur}, S.~M., {Mahzooni}, S., \& {Smitka},
  M.~T. 2015, \aj, 149, 117

\bibitem[{{Macri} {et~al.}(2006){Macri}, {Stanek}, {Bersier}, {Greenhill}, \&
  {Reid}}]{Macri06}
{Macri}, L.~M., {Stanek}, K.~Z., {Bersier}, D., {Greenhill}, L.~J., \& {Reid},
  M.~J. 2006, \apj, 652, 1133

\bibitem[{{Madore}(1982)}]{Madore82}
{Madore}, B.~F. 1982, \apj, 253, 575

\bibitem[{{Matsunaga} {et~al.}(2009){Matsunaga}, {Kawadu}, {Nishiyama},
  {Nagayama}, {Hatano}, {Tamura}, {Glass}, \& {Nagata}}]{Matsunaga09}
{Matsunaga}, N., {Kawadu}, T., {Nishiyama}, S., {et~al.} 2009, \mnras, 399,
  1709

\bibitem[{{Mouhcine} \& {Lan{\c c}on}(2003)}]{Mouhcine03}
{Mouhcine}, M., \& {Lan{\c c}on}, A. 2003, \mnras, 338, 572

\bibitem[{{Nicholls} {et~al.}(2009){Nicholls}, {Wood}, {Cioni}, \&
  {Soszy{\'n}ski}}]{Nicholls09}
{Nicholls}, C.~P., {Wood}, P.~R., {Cioni}, M.-R.~L., \& {Soszy{\'n}ski}, I.
  2009, \mnras, 399, 2063

\bibitem[{{Olivier} {et~al.}(2001){Olivier}, {Whitelock}, \&
  {Marang}}]{Olivier01}
{Olivier}, E.~A., {Whitelock}, P., \& {Marang}, F. 2001, \mnras, 326, 490

\bibitem[{{Payne-Gaposchkin}(1954)}]{PayneGaposchkin54}
{Payne-Gaposchkin}, C. 1954, Annals of Harvard College Observatory, 113, 189

\bibitem[{{Percy} {et~al.}(2004){Percy}, {Bakos}, {Besla}, {Hou}, {Velocci}, \&
  {Henry}}]{Percy04}
{Percy}, J.~R., {Bakos}, A.~G., {Besla}, G., {et~al.} 2004, in Astronomical
  Society of the Pacific Conference Series, Vol. 310, IAU Colloq. 193: Variable
  Stars in the Local Group, ed. D.~W. {Kurtz} \& K.~R. {Pollard}, 348

\bibitem[{{Percy} \& {Deibert}(2016)}]{Percy16}
{Percy}, J.~R., \& {Deibert}, E. 2016, Journal of the American Association of
  Variable Star Observers (JAAVSO), 44, 94

\bibitem[{{Planck Collaboration} {et~al.}(2016){Planck Collaboration}, {Ade},
  {Aghanim}, {Arnaud}, {Ashdown}, {Aumont}, {Baccigalupi}, {Banday},
  {Barreiro}, {Bartlett}, \& et~al.}]{Planck16XIII}
{Planck Collaboration}, {Ade}, P.~A.~R., {Aghanim}, N., {et~al.} 2016, \aap,
  594, A13

\bibitem[{{Poleski} {et~al.}(2010{\natexlab{a}}){Poleski}, {Soszy{\'n}ski},
  {Udalski}, {Szyma{\'n}ski}, {Kubiak}, {Pietrzy{\'n}ski}, {Wyrzykowski}, \&
  {Ulaczyk}}]{Poleski10b}
{Poleski}, R., {Soszy{\'n}ski}, I., {Udalski}, A., {et~al.} 2010{\natexlab{a}},
  \actaa, 60, 179

\bibitem[{{Poleski} {et~al.}(2010{\natexlab{b}}){Poleski}, {Soszy{\'n}ski},
  {Udalski}, {Szyma{\'n}ski}, {Kubiak}, {Pietrzy{\'n}ski}, {Wyrzykowski},
  {Szewczyk}, \& {Ulaczyk}}]{Poleski10a}
---. 2010{\natexlab{b}}, \actaa, 60, 1

\bibitem[{{Rejkuba}(2004)}]{Rejkuba04}
{Rejkuba}, M. 2004, \aap, 413, 903

\bibitem[{{Riess} {et~al.}(2016){Riess}, {Macri}, {Hoffmann}, {Scolnic},
  {Casertano}, {Filippenko}, {Tucker}, {Reid}, {Jones}, {Silverman},
  {Chornock}, {Challis}, {Yuan}, {Brown}, \& {Foley}}]{Riess16}
{Riess}, A.~G., {Macri}, L.~M., {Hoffmann}, S.~L., {et~al.} 2016, \apj, 826, 56

\bibitem[{{Riess} {et~al.}(2018){Riess}, {Casertano}, {Yuan}, {Macri},
  {Anderson}, {Mackenty}, {Bowers}, {Clubb}, {Filippenko}, {Jones}, \&
  {Tucker}}]{Riess18}
{Riess}, A.~G., {Casertano}, S., {Yuan}, W., {et~al.} 2018, ArXiv e-prints,
  arXiv:1801.01120

\bibitem[{{Rigault} {et~al.}(2015){Rigault}, {Aldering}, {Kowalski}, {Copin},
  {Antilogus}, {Aragon}, {Bailey}, {Baltay}, {Baugh}, {Bongard}, {Boone},
  {Buton}, {Chen}, {Chotard}, {Fakhouri}, {Feindt}, {Fagrelius}, {Fleury},
  {Fouchez}, {Gangler}, {Hayden}, {Kim}, {Leget}, {Lombardo}, {Nordin}, {Pain},
  {Pecontal}, {Pereira}, {Perlmutter}, {Rabinowitz}, {Runge}, {Rubin},
  {Saunders}, {Smadja}, {Sofiatti}, {Suzuki}, {Tao}, \& {Weaver}}]{Rigault15}
{Rigault}, M., {Aldering}, G., {Kowalski}, M., {et~al.} 2015, \apj, 802, 20

\bibitem[{{Saio} {et~al.}(2015){Saio}, {Wood}, {Takayama}, \& {Ita}}]{Saio15}
{Saio}, H., {Wood}, P.~R., {Takayama}, M., \& {Ita}, Y. 2015, \mnras, 452, 3863

\bibitem[{{Sesar} {et~al.}(2010){Sesar}, {Ivezi{\'c}}, {Grammer}, {Morgan},
  {Becker}, {Juri{\'c}}, {De Lee}, {Annis}, {Beers}, {Fan}, {Lupton}, {Gunn},
  {Knapp}, {Jiang}, {Jester}, {Johnston}, \& {Lampeitl}}]{Sesar10}
{Sesar}, B., {Ivezi{\'c}}, {\v Z}., {Grammer}, S.~H., {et~al.} 2010, \apj, 708,
  717

\bibitem[{{Soszy{\'n}ski}(2007)}]{Soszynski07b}
{Soszy{\'n}ski}, I. 2007, \apj, 660, 1486

\bibitem[{{Soszynski} {et~al.}(2005){Soszynski}, {Udalski}, {Kubiak},
  {Szymanski}, {Pietrzynski}, {Zebrun}, {Szewczyk}, {Wyrzykowski}, \&
  {Ulaczyk}}]{Soszynski05}
{Soszynski}, I., {Udalski}, A., {Kubiak}, M., {et~al.} 2005, \actaa, 55, 331

\bibitem[{{Soszynski} {et~al.}(2008){Soszynski}, {Poleski}, {Udalski},
  {Szymanski}, {Kubiak}, {Pietrzynski}, {Wyrzykowski}, {Szewczyk}, \&
  {Ulaczyk}}]{Soszynski08a}
{Soszynski}, I., {Poleski}, R., {Udalski}, A., {et~al.} 2008, \actaa, 58, 163

\bibitem[{{Soszy{\'n}ski} {et~al.}(2008){Soszy{\'n}ski}, {Udalski},
  {Szyma{\'n}ski}, {Kubiak}, {Pietrzy{\'n}ski}, {Wyrzykowski}, {Szewczyk},
  {Ulaczyk}, \& {Poleski}}]{Soszynski08b}
{Soszy{\'n}ski}, I., {Udalski}, A., {Szyma{\'n}ski}, M.~K., {et~al.} 2008,
  \actaa, 58, 293

\bibitem[{{Soszy{\'n}ski} {et~al.}(2009{\natexlab{a}}){Soszy{\'n}ski},
  {Udalski}, {Szyma{\'n}ski}, {Kubiak}, {Pietrzy{\'n}ski}, {Wyrzykowski},
  {Szewczyk}, {Ulaczyk}, \& {Poleski}}]{Soszynski09a}
---. 2009{\natexlab{a}}, \actaa, 59, 1

\bibitem[{{Soszy{\'n}ski} {et~al.}(2009{\natexlab{b}}){Soszy{\'n}ski},
  {Udalski}, {Szyma{\'n}ski}, {Kubiak}, {Pietrzy{\'n}ski}, {Wyrzykowski},
  {Szewczyk}, {Ulaczyk}, \& {Poleski}}]{Soszynski09b}
---. 2009{\natexlab{b}}, \actaa, 59, 239

\bibitem[{{Soszy{\'n}ski} {et~al.}(2009{\natexlab{c}}){Soszy{\'n}ski},
  {Udalski}, {Szyma{\'n}ski}, {Kubiak}, {Pietrzy{\'n}ski}, {Wyrzykowski},
  {Szewczyk}, {Ulaczyk}, \& {Poleski}}]{Soszynski09c}
---. 2009{\natexlab{c}}, \actaa, 59, 335

\bibitem[{{Soszy{\'n}ski} {et~al.}(2011){Soszy{\'n}ski}, {Udalski},
  {Szyma{\'n}ski}, {Kubiak}, {Pietrzy{\'n}ski}, {Wyrzykowski}, {Ulaczyk},
  {Poleski}, {Koz{\l}owski}, \& {Pietrukowicz}}]{Soszynski11}
---. 2011, \actaa, 61, 217

\bibitem[{{Soszy{\'n}ski} {et~al.}(2013){Soszy{\'n}ski}, {Udalski},
  {Szyma{\'n}ski}, {Kubiak}, {Pietrzy{\'n}ski}, {Wyrzykowski}, {Ulaczyk},
  {Poleski}, {Koz{\l}owski}, {Pietrukowicz}, \& {Skowron}}]{Soszynski13}
---. 2013, \actaa, 63, 21

\bibitem[{{Stetson}(1987)}]{Stetson87}
{Stetson}, P.~B. 1987, \pasp, 99, 191

\bibitem[{{Stetson}(1994)}]{Stetson94}
---. 1994, \pasp, 106, 250

\bibitem[{{Stetson}(1996)}]{Stetson96}
---. 1996, \pasp, 108, 851

\bibitem[{{STScI Development Team}(2013)}]{PySynPhot}
{STScI Development Team}. 2013, {pysynphot: Synthetic photometry software
  package}, Astrophysics Source Code Library, ascl:1303.023

\bibitem[{{Trabucchi} {et~al.}(2017){Trabucchi}, {Wood}, {Montalb{\'a}n},
  {Marigo}, {Pastorelli}, \& {Girardi}}]{Trabucchi17}
{Trabucchi}, M., {Wood}, P.~R., {Montalb{\'a}n}, J., {et~al.} 2017, \apj, 847,
  139

\bibitem[{{Udalski} {et~al.}(2008){Udalski}, {Szymanski}, {Soszynski}, \&
  {Poleski}}]{Udalski08}
{Udalski}, A., {Szymanski}, M.~K., {Soszynski}, I., \& {Poleski}, R. 2008,
  \actaa, 58, 69

\bibitem[{{Whitelock} {et~al.}(1994){Whitelock}, {Menzies}, {Feast}, {Marang},
  {Carter}, {Roberts}, {Catchpole}, \& {Chapman}}]{Whitelock94}
{Whitelock}, P., {Menzies}, J., {Feast}, M., {et~al.} 1994, \mnras, 267, 711

\bibitem[{{Whitelock} {et~al.}(2006){Whitelock}, {Feast}, {Marang}, \&
  {Groenewegen}}]{Whitelock06}
{Whitelock}, P.~A., {Feast}, M.~W., {Marang}, F., \& {Groenewegen}, M.~A.~T.
  2006, \mnras, 369, 751

\bibitem[{{Whitelock} {et~al.}(2008){Whitelock}, {Feast}, \& {Van
  Leeuwen}}]{Whitelock08}
{Whitelock}, P.~A., {Feast}, M.~W., \& {Van Leeuwen}, F. 2008, \mnras, 386, 313

\bibitem[{{Whitelock} {et~al.}(2003){Whitelock}, {Feast}, {van Loon}, \&
  {Zijlstra}}]{Whitelock03}
{Whitelock}, P.~A., {Feast}, M.~W., {van Loon}, J.~T., \& {Zijlstra}, A.~A.
  2003, \mnras, 342, 86

\bibitem[{{Whitelock} {et~al.}(2017){Whitelock}, {Kasliwal}, \&
  {Boyer}}]{Whitelock17}
{Whitelock}, P.~A., {Kasliwal}, M., \& {Boyer}, M. 2017, in European Physical
  Journal Web of Conferences, Vol. 152, European Physical Journal Web of
  Conferences, 01009

\bibitem[{{Whitelock} {et~al.}(2009){Whitelock}, {Menzies}, {Feast},
  {Matsunaga}, {Tanab{\'e}}, \& {Ita}}]{Whitelock09}
{Whitelock}, P.~A., {Menzies}, J.~W., {Feast}, M.~W., {et~al.} 2009, \mnras,
  394, 795

\bibitem[{{Wood} {et~al.}(1999){Wood}, {Alcock}, {Allsman}, {Alves}, {Axelrod},
  {Becker}, {Bennett}, {Cook}, {Drake}, {Freeman}, {Griest}, {King}, {Lehner},
  {Marshall}, {Minniti}, {Peterson}, {Pratt}, {Quinn}, {Stubbs}, {Sutherland},
  {Tomaney}, {Vandehei}, \& {Welch}}]{Wood99}
{Wood}, P.~R., {Alcock}, C., {Allsman}, R.~A., {et~al.} 1999, in IAU Symposium,
  Vol. 191, Asymptotic Giant Branch Stars, ed. T.~{Le Bertre}, A.~{Lebre}, \&
  C.~{Waelkens}, 151

\bibitem[{{Yoachim} {et~al.}(2009){Yoachim}, {McCommas}, {Dalcanton}, \&
  {Williams}}]{Yoachim09}
{Yoachim}, P., {McCommas}, L.~P., {Dalcanton}, J.~J., \& {Williams}, B.~F.
  2009, \aj, 137, 4697

\bibitem[{{Yuan} {et~al.}(2017{\natexlab{a}}){Yuan}, {He}, {Macri}, {Long}, \&
  {Huang}}]{Yuan17a}
{Yuan}, W., {He}, S., {Macri}, L.~M., {Long}, J., \& {Huang}, J.~Z.
  2017{\natexlab{a}}, \aj, 153, 170

\bibitem[{{Yuan} {et~al.}(2017{\natexlab{b}}){Yuan}, {Macri}, {He}, {Huang},
  {Kanbur}, \& {Ngeow}}]{Yuan17b}
{Yuan}, W., {Macri}, L.~M., {He}, S., {et~al.} 2017{\natexlab{b}}, ArXiv
  e-prints, arXiv:1708.04742

\end{thebibliography}

\end{document}